\documentclass[10pt,A4paper,conference]{IEEEtran}
\IEEEoverridecommandlockouts
\usepackage{cite}
\usepackage{amsmath,amssymb,amsfonts}
\usepackage{graphicx}
\usepackage{textcomp}
\usepackage{xcolor}
\usepackage{booktabs}
\usepackage{tabularx}
\usepackage{color, colortbl}
\usepackage[misc]{ifsym}
\usepackage{xcolor}
\definecolor{lime}{rgb}{0.88,2,10}

\renewcommand{\baselinestretch}{.982}
\usepackage{subcaption}
\usepackage{wrapfig}
\usepackage{hyperref}
\usepackage[linesnumbered,ruled,vlined]{algorithm2e}
\usepackage{xpatch}
\usepackage{url}
\usepackage{multirow}
\usepackage{multicol}
\usepackage{wrapfig}
\usepackage[ruled,vlined,linesnumbered]{algorithm2e}

\usepackage{subcaption} 
\usepackage[export]{adjustbox}
\usepackage{nomencl}
\usepackage{siunitx}
\usepackage{blindtext}
\usepackage[T1]{fontenc}
\usepackage[spaces,hyphens]{xurl}
\usepackage{float}
\usepackage{graphicx}
\usepackage{colortbl}

\newcommand{\fref}[1]{Fig.~\ref{#1}}

\usepackage[misc]{ifsym}
\usepackage{multirow}
\usepackage{enumitem}
\newcommand*{\Resize}[2]{\resizebox{#1}{!}{$#2$}}%

\SetAlFnt{\small}

\SetKwComment{Comment}{$\triangleright$\ }{}

\usepackage[utf8]{inputenc}
\usepackage{enumitem}

\SetKwInput{kwInit}{Initialize}

\def\BibTeX{{\rm B\kern-.05em{\sc i\kern-.025em b}\kern-.08em
    T\kern-.1667em\lower.7ex\hbox{E}\kern-.125emX}}

\usepackage[numbers,sort&compress]{natbib}
\usepackage[font={small}]{caption} 

\usepackage[misc]{ifsym}

\usepackage{tikz}

\newcounter{stepnum}

\usepackage[outline]{contour}
\contourlength{0.2pt}
\contournumber{15}

\newcommand\HUGE{\fontsize{19.9}{25}\selectfont}

\usepackage{fancyhdr}
\fancypagestyle{mystyle}{
    \chead{\large The 22nd IEEE International Conference on Mobile Ad-Hoc and Smart Systems (MASS 2025)}}

\begin{document}
\title{\HUGE {Detecting Untargeted Attacks and Mitigating Unreliable Updates in Federated Learning for Underground Mining Operations}}

\author{\IEEEauthorblockN{
 Md Sazedur Rahman\IEEEauthorrefmark{1}, Mohamed Elmahallawy\IEEEauthorrefmark{2}, Sanjay Madria\IEEEauthorrefmark{1}, Samuel Frimpong\IEEEauthorrefmark{3}}  
      \IEEEauthorblockA{%
 \IEEEauthorrefmark{1}Computer Science Department, Missouri University of Science and Technology, Rolla, MO 65401, USA}
       \IEEEauthorblockA{%
 \IEEEauthorrefmark{2}School of Engineering \& Applied Science,  Cybersecurity Program, Washington State University, Richland, WA 99354, USA}
   \IEEEauthorblockA{%
 \IEEEauthorrefmark{3}Explosive \& Mining
Engineering Department, Missouri University of Science and Technology, Rolla, MO 65401, USA}
\thanks{\IEEEauthorrefmark{4}This work was supported by a grant from CDC-NIOSH.} 
Emails:  {mrvfw@mst.edu, mohamed.elmahallawy@wsu.edu, madrias@mst.edu, frimpong@mst.edu}\vspace{-0.15cm}}
    

\maketitle
\thispagestyle{mystyle}
\begin{abstract}
Underground mining operations rely on distributed sensor networks to collect critical data daily, including mine temperature, toxic gas concentrations, and miner movements for hazard detection and operational decision-making. However, transmitting raw sensor data to a central server for training deep learning models introduces significant privacy risks, potentially exposing sensitive mine-specific information. Federated Learning (FL) offers a transformative solution by enabling collaborative model training while ensuring that raw data remains localized at each mine. Despite its advantages, FL in underground mining faces key challenges: (i) An attacker may compromise a mine's local model by employing techniques such as sign-flipping attacks or additive noise, leading to erroneous predictions; (ii) Low-quality (yet potentially valuable) data, caused by poor lighting conditions or sensor inaccuracies in mines may degrade the FL training process. In response, this paper proposes MineDetect, a {\em defense} FL framework that detects and isolates the attacked models while mitigating the impact of mines with low-quality data. MineDetect introduces two key innovations: (i) Detecting attacked models (maliciously manipulated) by developing a history-aware mechanism that leverages local and global averages of gradient updates; (ii) Identifying and eliminating adversarial influences from unreliable models (generated by clients with poor data quality) on the FL training process. Comprehensive simulations across diverse datasets demonstrate that MineDetect outperforms existing methods in both robustness and accuracy, even in challenging non-IID data scenarios. Its ability to counter adversarial influences while maintaining lower computational efficiency makes it a vital advancement for improving safety and operational effectiveness in underground mining. Our code is available at: \href{https://github.com/sazed49/IEEEMASS2025}{https://github.com/sazed49/IEEEMASS2025}

\end{abstract}
\section{Introduction}
Underground mining operations depend on distributed sensor networks to gather extensive data, like mine temperatures, concentrations of hazardous gases, and the movement of miners which are essential to ensure safety and enabling real-time decision-making \cite{10723487,yadav2025predicting}. However, transmitting raw sensor data to a central server for training large-scale deep learning (DL) models introduces significant security risks.
Federated Learning (FL) \cite{mcmahan2017communication}  allows mines to collaboratively train DL models without sharing their raw data. Specifically, each mine utilizes its collected sensor data or video surveillance footage to locally train a DL model for various classification tasks \cite{wu2024lightweight}. These locally trained models are then sent to a parameter server $\mathcal{PS}$ for aggregation, generating a global model that enhances prediction accuracy among all mines.

{\bf Challenges.} Despite the numerous benefits of integrating FL into underground mining, its collaborative nature inherently exposes it to a range of potential attacks, such as targeted and untargeted attacks \cite{lyu2022privacy,wei2023client}. In {\em untargeted attacks} (e.g., sign-flipping or additive-noise), the attacker may manipulate its generated local parameters, thereby undermining the overall performance of the FL global model \cite{cao2020fltrust,wu2020federated}. In \emph{sign-flipping attacks}, a compromised mine intentionally reverses the signs of its parameter and sends them to the central server 
which can mislead the global model into interpreting increasing seismic risks as decreasing ones. Similarly, in an \emph{additive noise attack}, a compromised mine injects random noise into the parameter updates generated from toxic gas sensor data which can mislead the global model to accurately predict hazardous gas levels. In {\em targeted attacks} (e.g, label-flipping or multi-label-flipping), the adversary focuses on degrading the global model's performance for specific data types or tasks while leaving other functionalities unaffected \cite{gupta2022long,yin2023defending}. 

On the other hand, an \emph{unreliable mine}, which collects low-quality or noisy data due to challenging environmental conditions such as extreme darkness that impact images captured by surveillance cameras \cite{jewel2024dis}, resulting in hindering of global model to convergence. Despite this, {\em we cannot ignore the models of unreliable mines in the aggregation process, as they introduce diversity into the dataset, which can be valuable for capturing rare conditions}.
Among the three risks to FL in mines—untargeted attacks, targeted attacks, and unreliable mines--the targeted attack is easier to detect through parameter pattern analysis \cite{yazdinejad2024robust,sun2023attacking}. For untargeted attacks, the work in \cite{lewis2024mitigation} proposed a scheme called SecAdam, which uses a client-side adaptive optimizer to counteract sign-flipping attacks, along with a secure aggregation technique that accurately aggregates updates from the client-side adaptive optimizer. However, it focuses solely on one type of attack, which is impractical, as compromised clients in real-world scenarios may execute multiple attack types. 
Another approach, proposed by Gupta et al. \cite{gupta2022long}, introduces an algorithm called MUD-HoG, which utilizes a long-short history of parameters to differentiate between various attack types and unreliable clients. A significant drawback of this method is the reliance on parameters from the past three rounds to create a short history for detecting untargeted attacks. In practical scenarios, clients may behave normally in some rounds and act maliciously in others, further complicating detection.

{\bf Contributions.} In response, this paper proposes {\em MineDetect}, a robust FL framework specifically designed for an underground mining environment. MineDetect detects untargeted attacks on mines and mitigates the negative effects of unreliable mines by analyzing their parameter updates.  Specifically, MineDetect incorporates two key innovations: (i) Local average that computes the average of a mine's current parameter/weight update and its previous-round update, capturing temporal trends to detect malicious behaviors; (ii) Global average that calculates the mean of all mines' local averages, serving as a robust benchmark for identifying compromised and unreliable mines. In  summary, this paper makes the following contributions:
\begin{itemize}[leftmargin=*]
    \item  We propose MineDetect, a history-aware FL framework designed to defend against untargeted attacks in underground mining while mitigating the impact of unreliable mines with low-quality data on the global model.

   \item MineDetect's defense scheme detects {\em mixed types of attackers} by (i) utilizing cosine similarity to identify sign-flipping attacks (Algorithm~\ref{algorithm2}), (ii) applying variance analysis over historical parameters to detect additive noise attacks (Algorithm~\ref{algorithm3}), and (iii) distinguishing between untargeted attacks and unreliable mines using Euclidean distance deviations (Algorithm~\ref{algorithm4})---ensuring the robustness of the global model.  
 
\item We comprehensively evaluate MineDetect under various settings, including both IID and non-IID scenarios, which are more realistic and challenging. Our evaluation utilizes diverse datasets, namely the Rockburst classification dataset (a real underground mining dataset), MNIST, and Fashion-MNIST. The results demonstrate that MineDetect consistently outperforms state-of-the-art techniques in terms of robustness, accuracy, and efficiency. Specifically, MineDetect enhances accuracy by 5.94\% comparing with best performing baseline method and decreases \textit{false positive rate} in adversarial conditions by 8.2\%.
\end{itemize}

\section{Related Work}\label{sec:related works}
FL global model aggregation is typically divided into two main categories: (1) {\em Cryptographic-based methods} such as multi-party computation~\cite{zhou2024secure}, homomorphic encryption (HE)~\cite{pan2024fedshe}, and functional encryption~\cite{zhang2024efficient}; (2) {\em Non-cryptographic based aggregation methods}, such as Krum~ and Multi-Krum~\cite{blanchard2017machine} to isolate compromised or malicious clients before aggregating them into the global model. Below is a brief overview of each of these approaches.

\subsection{Cryptographic-Based Aggregation Methods}
The authors of \cite{zhang2023privacyeafl} have proposed a privacy-enhanced aggregation method named PrivacyEAFL for FL in mobile crowdsensing by avoiding collusion and data leakage using hashed Diffie-Hellman (DH) key exchange and a two-trapdoor homomorphic cryptosystem. Clients and the server establish secure pairwise keys using DH.
The biggest drawback is the higher processing burden on client devices due to sophisticated cryptographic processes. Recently another work \cite{pan2024fedshe} has presented FedSHE, a safe framework based on adaptive segmented CKKS homomorphic encryption to resist parameter leaking attacks. It has proposed a depth-adaptive key-generating mechanism to maximize security settings. In FL environments, FedSHE offers stronger defence against privacy invasions and more efficiency than tools like Paillier. The authors of \cite{zhang2024efficient} have presented PIM-MCFE, a multi-client functional encryption system grounded on the Learning with Errors (LWE) assumption for a privacy-preserving federated learning (PPFL) framework. The high storage demand resulting from big key sizes needed for LWE-based encryption limits scalability in resource-limited surroundings.

\subsection{ Non-Cryptographic Based Aggregation Methods}
\label{sec:non-crypto}
 One of the first and most widely known malicious resilient aggregation methods is Krum \cite{blanchard2017machine}, which identifies the most representative parameter by minimizing its distance to the majority of the other updates.  A variant of Krum, called Multi-Krum \cite{blanchard2017machine}, proposed on the same Krum paper, attempts to balance convergence speed with robustness, but both are only useful for IID scenarios. In the same year, the authors of \cite{chen2017distributed} have introduced GeoMed, which uses geometric median aggregation to improve robustness in FL. A significant constraint is its sensitivity to batch size. Another work by Fung et al. \cite{fung2020limitations} called FoolsGold detectS and mitigateS Sybil clients attempting to manipulate the global model through coordinated adversarial updates. However, FoolsGold is limited in its effectiveness when Sybil attackers employ highly adaptive strategies. Another approach, called MUD-Hog \cite{gupta2022long}, proposed a detection scheme that utilizes historical parameter information to differentiate between malicious and legitimate clients. However, MUD-Hog’s reliance on a brief history of parameters—limited to the last three rounds--poses challenges in detecting attacks early. 
Finally, attestedFL \cite{al2023untargeted}  tracks client behavior using metrics like angular distance and convergence speed to detect malicious updates. But,its high computational costs make it less scalable.
In this work, we propose MineDetect, a non-cryptographic-based approach for detecting sign-flipping and additive noise attacks, as well as addressing unreliable clients, all while assuring better accuracy compared to existing approaches in the literature.

\section{MineDetect's System Model and Threat Model}
MineDetect is a history-aware FL framework that protects the global  detection model in underground mining from various malicious attacks and unreliable mine behaviors. MineDetect's system model considers the following types of mines:
\begin{itemize}[leftmargin=*]
    \item {\bf Normal/Benign Mines:} These mines honestly participate in the FL training process by locally training their models and sharing the updates with $\mathcal{PS}$ without any manipulation. 
    
    \item {\bf Compromised Mines:} These mines intentionally disrupt the FL training process using mixed attack types. For instance, they may reverse the sign of model updates or introduce random noise before sending their models to the $\mathcal{PS}$, aiming to degrade the global model's accuracy.

    \item {\bf Unreliable Mines:} These mines honestly participate in FL but provide low-quality data, which, while offering valuable diversity, negatively impacts the global model’s accuracy.
 
\end{itemize}

\subsection{MineDetect's System Model}

\begin{figure*}[!t]
\centering
\includegraphics[width=0.8\linewidth]{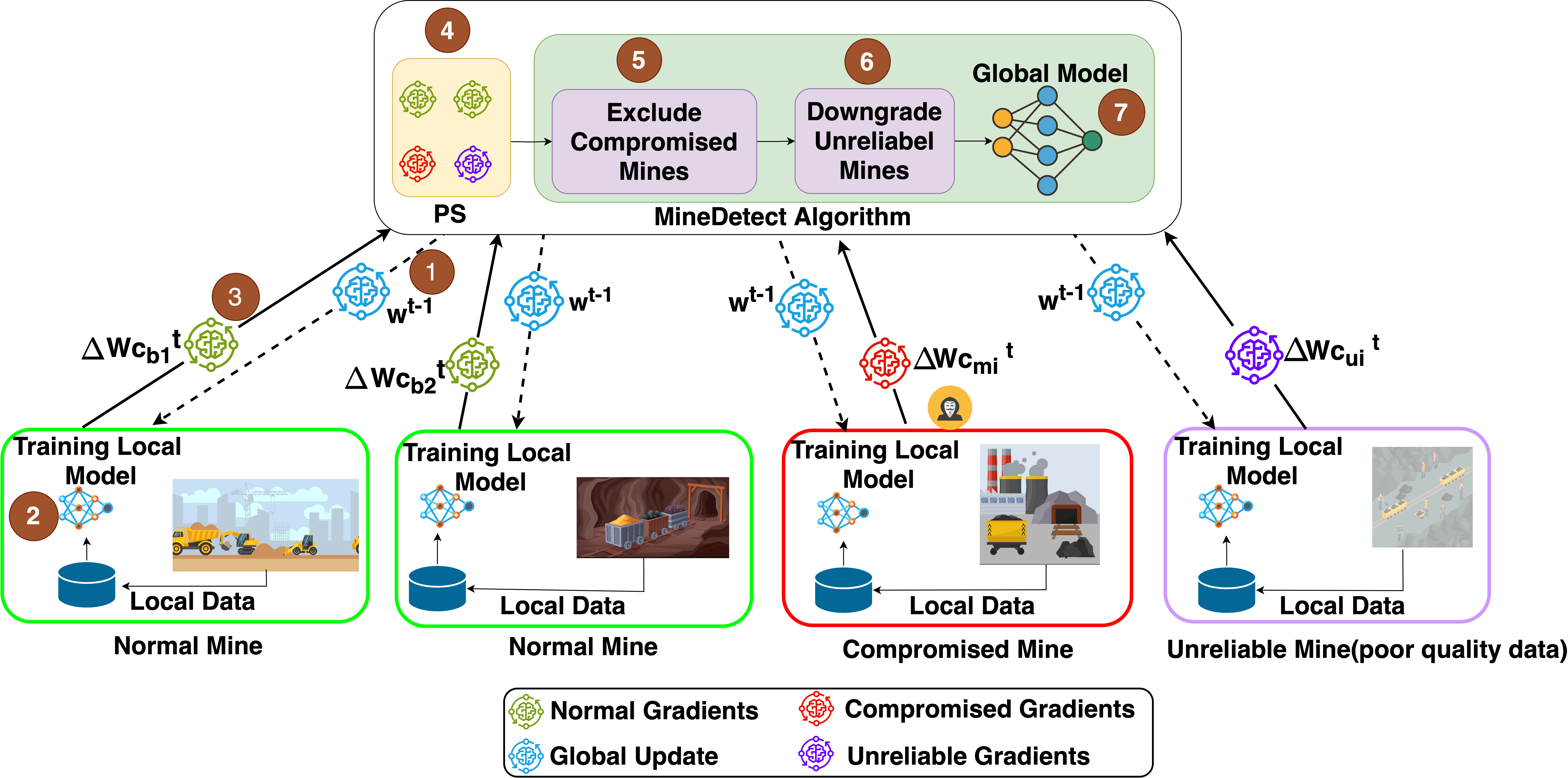}
\caption{Overview of the FL framework in underground mining with compromised and unreliable updates.} \vspace{-4mm}
\label{fig:figure1} 
\end{figure*}

We consider a realistic FL system with three distinct types of clients (mines) as illustrated in \fref{fig:figure1}. Let $\mathcal{N} = \{\mathcal{C}_{B}\cup  \mathcal{C}_{M}\cup \mathcal{C}_{U}\}$ be the set of mines participating in the FL training process, where  $\mathcal{C}_{B} = \{C_{b_1}, C_{b_2}, \dots, C_{B}\}$ represents $B$ normal/benign mines that follow standard FL training procedures, $\mathcal{C}_{M} = \{C_{m_1}, C_{m_2}, \dots, C_{M}\}$ represents $M$ compromised/malicious mines $(M \ll N)$ that launch sign-flipping or additive noise attacks, $\mathcal{C}_{U} = \{C_{u_1}, C_{u_2}, \dots, C_{U}\}$ represents $U$ unreliable mines $(U \ll N)$. All mines share a common model structure, working toward the same learning objective. The FL training process begins with the (\(\mathcal{PS}\)) distributing the global model \( \boldsymbol{w}^{t-1} \) to all participating mines. Then, each mine \(C_i\in \mathcal{N}\) utilizes \( \boldsymbol{w}^{t-1} \) and its local data to updates its local model parameters \( \Delta \boldsymbol{w}_{C_i}^t \) by minimizing a cross-entropy loss function \( \mathcal {L}(h_{\boldsymbol{w}_{C_i}}(x), y) \) over a set number of local epochs, where \( h_{\boldsymbol{w}_{C_i}}(.) \) is the hypothesis function predicting \( y \) for any input \( x \) by a mine $C_i$. Upon completing local training, each mine \( C_i\) transmits its updated parameters \( \Delta \boldsymbol{w}_{C_i}^t \) back to the \(\mathcal{PS}\), where \( \Delta \boldsymbol {w}_{C_i}^t \) represents the {\em gradient} updates of the parameters. However, each compromised mine \( C_{m_i} \)  manipulates its local parameters and sends adversarial gradients updates \( \Delta \boldsymbol{w}_{C_{m_i}}^t \). Similarly,  each unreliable mine \( C_{u_i} \) generates parameter updates  \( \Delta \boldsymbol{w}_{C_{u_i}}^t \) with significant variance compared to the updates from normal mines. Mathematically, this training process can be expressed as:

\begin{equation}
\Delta \boldsymbol{w}_{C_i}^t = \boldsymbol{w}^{t-1} - \arg\min_{\boldsymbol{w}} \mathcal{L}(h_{\boldsymbol{w}_{C_i}}(x), y)
\end{equation}
If each mine $C_i$ operates as a benign mine, then the server combines all the local gradients as:
\begin{equation}\label{eq:2}
\boldsymbol{w}^t(normal) = \sum_{C_i\in \mathcal{N}} \frac{|D_{c_i}|}{|D|} \Delta \boldsymbol{w}_{C_i}^t
\end{equation}
where \(|D_{C_i}|\) is the dataset size owned by a mine \(C_i\) and \(|D|\) is total amount of data among all mines. Finally, the updated gradient of the server for the next round becomes:
\begin{equation}\label{eq:3}
\boldsymbol{w}^{t+1} = \boldsymbol{w}^{t-1} - \eta \boldsymbol{w}^t
\end{equation}
where \(\eta\) is the learning rate.

\subsection{Problem Definition}
The primary goal of the $\mathcal{PS}$ is to accurately {\em detect and isolate compromised mines from contributing to the global model, while mitigating the negative impact of unreliable mines on convergence during each global training round}.  So, the aggregation equation \eqref{eq:2} should be modified to prioritize updates from normal/benign mines while down-weighting unreliable mines. This can be mathematically expressed as:
\begin{align}
\boldsymbol {w}^t (normal+unreliable)&=\sum_{C_{b_i} \in \mathcal{C}_{B}} \frac{|D_{C_{b_i}}|}{|D|} \Delta \boldsymbol{w}_{C_i}^t\\ &+ \beta \sum_{C_{u_i} \in \mathcal{C}_{U}} \frac{|D_{C_{u_i}}|}{|D|} \Delta \boldsymbol{w}_{C_{u_i}}^t\nonumber
\end{align}

\noindent where \( \boldsymbol {w}^t (normal+unreliable) \) is the aggregated model update at round \( t \) considering both sets of normal mines  $\mathcal{C}_{B}$ and unreliable mines $\mathcal{C}_{U}$. \( |D_{C_{b_i}}| \) and \( |D_{C_{u_i}}| \) denote the number of data samples owned by normal mine \( C_{b_i} \) and unreliable mine \( C_{b_i} \), respectively, while \( \Delta \boldsymbol{w}_{C_{b_i}}^t \) and \( \Delta \boldsymbol{w}_{C_{ui}}^t \) are the local model update from normal mine \(C_{b_i} \)  and unreliable mine \( C_{u_i} \), respectively at round \( t \). The parameter \( \beta \) represents a weighting factor applied to the updates from unreliable mines to mitigate their effects on the global model. Thus, \eqref{eq:2} can be modified to obtain the updated gradient for the next round as:

\begin{equation}
\boldsymbol{w}^{t+1}(normal+unreliable) = \boldsymbol{w}^{t-1} - \eta \boldsymbol {w}^t 
\end{equation}

\subsection{MineDetect's Threat Model}

MineDetect considers that the $\mathcal{PS}$ is honest and solely responsible for identifying and excluding compromised mines. Additionally, it can downgrade the contributions of unreliable mines. It also considers that the total number of compromised mines is lower than the combined total of normal and unreliable mines. Moreover,no mine can access the models of other mines or collude with the $\mathcal{PS}$. A compromised mine can employ one of the following two types of untargeted attacks:

\begin{enumerate}[leftmargin=*]
    \item \emph{Sign Flipping Attack}: Any compromised mine reverses the direction of its gradient updates after completing its local training while keeping the magnitude unchanged. Mathematically, if the honest mine’s gradient update is \( \Delta \boldsymbol{w}_{C_i}^t \), a sign-flipping attacker generates $\Delta \boldsymbol{w}_{C_{m_i}}^t = -\Delta \boldsymbol{w}_{C_i}^t$, Where \( \Delta \boldsymbol{w}_{C_{m_i}}^t \) denotes the gradient updates in the opposite direction of the generated one.
    
    \item \emph{Additive Noise Attack}: A Compromised mine adds random noise  \( \alpha_{C_i}^t \sim \mathcal{N}(0, \sigma^2) \) to its gradient updates, altering the magnitude but not affecting the direction. Mathematically, if the honest mine’s gradient update is \( \Delta \boldsymbol{w}_{C_i}^t \), an additive noise attacker generates \( \Delta \boldsymbol{w}_{C_{m_i}}^t = \Delta \boldsymbol{w}_{C_i}^t + \boldsymbol{\alpha}_{C_i}^t \), where \( \boldsymbol{\alpha}_{C_i}^t \sim \mathcal{N}(\mathbf{0}, \sigma^2 \mathbf{I}) \) represents a vector of Gaussian noise added to the gradients of mine \(C_i\), drawn from a normal distribution with zero mean and variance \( \sigma^2 \).

\end{enumerate}

\section{MineDetect Framework}

The MineDetect framework comprises {\em seven} key steps, as illustrated in \fref{fig:figure1}, highlighted in circled brown. Steps 1–3 follow a standard FL process, where the \( \mathcal{PS} \) distributes the global model to all mines, each mine updates its local model, and then sends it back to the \( \mathcal{PS} \) for aggregation. However, MineDetect introduces crucial enhancements in Steps 4–7. It first examines them to exclude compromised mines and identify unreliable ones for down-weighting. Specifically, MineDetect performs two primary tasks: (1) Tracking gradients of all mines through {\em local and global averages}, and (2) Detecting and isolating compromised updates from malicious mines using statistical measures while mitigating the impact of unreliable updates. The effectiveness of MineDetect's detection mechanism relies on two key principles:
\begin{enumerate}[leftmargin=*]
\item {\bf Local Average.} The \textit{local average} serves as a historical reference for tracking the updated gradients of individual mines across training rounds in FL. For each mine \(C_i\), the local average is initialized with the initial model weights \(\Delta w_{C_i}^0\) in the first round, and in subsequent rounds, it is iteratively updated as:
\begin{equation}
\boldsymbol{L}_{C_i}^t = \frac{\boldsymbol{L}_{C_i}^{t-1} + \Delta \boldsymbol{w}_{C_i}^t}{2}, \quad \forall i \in \{1, 2, ..., N\}
\end{equation}
where \( \boldsymbol{L}_{C_i}^t \) denotes the local average at round \( t \) for mine \(C_i\), \( \Delta \boldsymbol{w}_{C_i}^t \) denotes the current model update of mine \( C_i \), and \( \boldsymbol{L}_{C_i}^{t-1} \) is the local average from the previous round. This formulation ensures that the local average reflects both the mine's current training performance and historical updates, providing a smooth, history-aware reference point.

\item {\bf Global Average.} The \textit{global average} is a centralized measure that aggregates local averages of all mines in each round to form a reliable global representation. It is computed as:
\begin{equation}
\boldsymbol{G}^t = \frac{1}{N} \sum\limits_{C_i\in \mathcal{N}} \boldsymbol{L}_{C_i}^t
\end{equation}
where \( \boldsymbol{G}_t \) is the global average at round \( t \). 
\end{enumerate}
\begin{algorithm}[!t]
\caption{MineDetect Algorithm}\label{algorithm1}
\KwIn{ 
Each mine \( C_i \) where \( C_i \in \mathcal{C}_{B} \cup \mathcal{C}_{M} \cup \mathcal{C}_{U} \), Update vectors \( \Delta \boldsymbol{w}_{C_i}^t \), global rounds \( T \), number of mines \( \mathcal{N} \)}
\KwOut{Set of Sign flip attackers \( S_{\text{flip}} \), Set of Additive noise attackers \( S_{\text{add}} \), Set of Unreliable mines \( S_{\text{unrl}} \)}
\For{\( t = 1 \) to \( T \)}{
\SetKwProg{Fn}{Function}{}{}
\Fn{Call Algorithm 2}{
    \KwRet{Sign flip attackers \( S_{\text{flip}} \)}
}

\SetKwProg{Fn}{Function}{}{}
\Fn{Call Algorithm 3}{
    \KwRet{Additive noise attackers \( S_{\text{add}} \)}
}

\SetKwProg{Fn}{Function}{}{}
\Fn{Call Algorithm 4}{
    \KwRet{Unreliable mines \( S_{\text{unrl}} \)}
}
 \( S = \mathcal{N} - S_{\text{flip}} - S_{\text{add}}  \)
        
        Aggregate parameters using   (4)\\
        Update global model \( \boldsymbol{w}^{t+1} \) using   (5)\\
        Send \( \boldsymbol{w}^{t+1} \) to each mine
    }
    \Return \( S_{\text{flip}}, S_{\text{add}}, S_{\text{unrl}} \)
\end{algorithm}
At each global round \( t \), Algorithm \ref{algorithm1} detects sign-flip attackers, additive noise attackers, and unreliable mines. It then aggregates updates only from a set \( S \), which includes both normal and unreliable mines to generate the new global model \( w^{t+1} \). Finally, this global model is distributed to all mines to start the next global round.

\subsection{Sign-Flipping Attack Detection scheme}
\begin{figure}[!t]
    \subfloat[Sign Flipping Attack Vector.]{
        \includegraphics[width=0.26\textwidth,height=2cm]{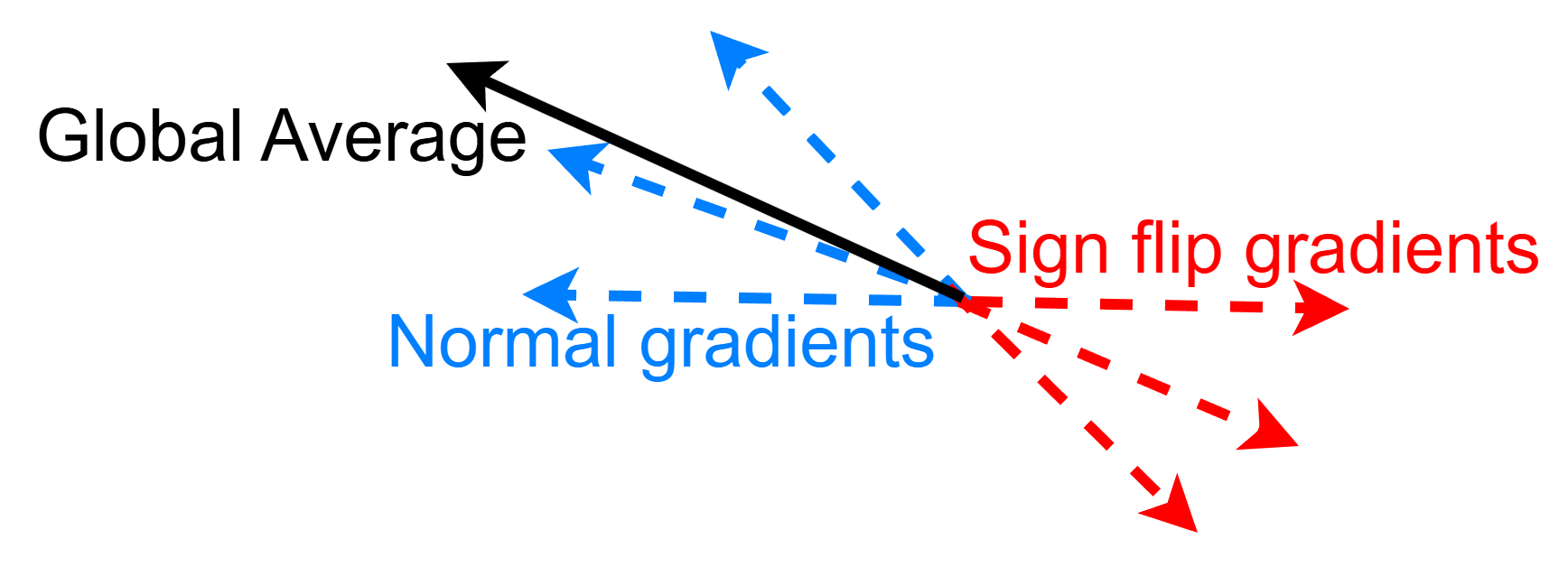}
        \label{fig:sign-flip}}
    \subfloat[Additive Noise Attack Vector.]{%
        \includegraphics[width=0.21\textwidth,height=2cm]{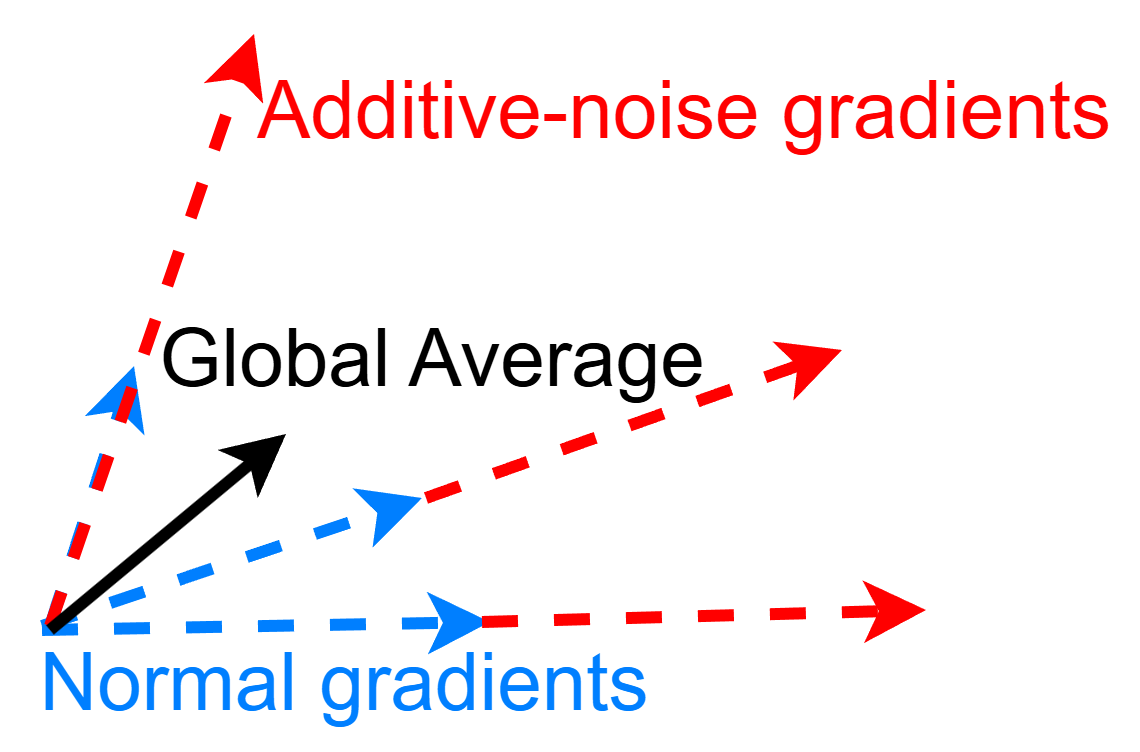}
        \label{fig:additive-noise}
    }
    \caption{An Illustration of untargeted attacks in MineDetect.  }
    \label{fig:attack-vectors}
\end{figure}

Compromised mines performing a sign-flip attack intentionally reverse the direction of their parameter updates to mislead the global model. \fref{fig:sign-flip} illustrates this phenomenon, where normal parameter updates (blue arrows) point in the general direction of the global average (black arrow), improving model convergence. In contrast, sign-flip updates (red arrows) are deliberately oriented in the opposite direction, aiming to sabotage learning. The detection mechanism relies on computing the {\bf cosine similarity} between each mine’s local average and the global average as follows:

\begin{equation}
d_{C_i} = \frac{\boldsymbol{L}_{C_i}^t \cdot \boldsymbol{G}^t}{\|\boldsymbol{L}_{C_i}^t\| \|\boldsymbol{G}^t\|}
\end{equation}
where \( d_{C_i} \) is the cosine similarity between mine \( C_i\)'s local average and the global average. Since the number of normal mines is significantly greater than the number of sign-flip attackers, the global average parameter naturally aligns with the majority of honest updates. As a result, any mine with an updated parameter direction opposite to the global trend (i.e., an angle greater than \(90^\circ\)) is highly likely to be a sign-flip attacker.

\begin{algorithm}[!t]
\caption{Sign Flipping Attack Detection Scheme}\label{algorithm2}
\KwIn{Each mine \( C_i \) where \( C_i \in \mathcal{C}_{B} \cup \mathcal{C}_{M} \cup \mathcal{C}_{U} \), Number of mines \( \mathcal{N} \)}
\KwOut{Set of sign flip attackers \( S_{\text{flip}} \)}

$S_{\text{flip}} \gets \emptyset$ \\
Compute local average \( \boldsymbol{L}_{C_i}^t \) for each mine \( C_i \) using Eq.~(6)\\
Compute global average \( \boldsymbol{G}^t \) using Eq.~(7) \\

\For{$i \gets 1$ \textbf{to} $\mathcal{N}$}{
      \( d_{C_i} \gets \frac{{\boldsymbol{L}}_{C_i}^t \cdot \boldsymbol{G}^t}{\|\boldsymbol{L}_{C_i}^t\| \|G_t\|}\)    \tcp{cosine similarity}

    \If{$d < 0$}{
        $S_{\text{flip}} \gets S_{\text{flip}} \cup \{C_i\}$  \tcp{Mark as sign flip attack}
    }
}

\Return $S_{\text{flip}}$
\end{algorithm}
Algorithm~\ref{algorithm2} describes how sign-flip attackers are detected by calculating the cosine similarity between each mine’s local average and the global average.


\subsection{Additive Noise Attack Detection Scheme}

Additive noise attackers introduce random Gaussian noise into their parameters without reversing the direction. Their parameter updates spread randomly, deviating from the global average and normal parameter distribution, as shown in \fref{fig:additive-noise}. MineDetect identifies such attackers by analyzing the historical variance of mines' local average across recent rounds and comparing their updated parameter magnitudes against a defined threshold. This is done using two detection approaches as follows:
\vspace{2mm}

{\noindent \bf A. Variance-Based Detection.}  The gradient updates exhibit abnormally high variations across consecutive training rounds. To quantify this, we track each mine’s historical local averages over the last five rounds, forming a history window:
\begin{equation}
\boldsymbol{H}_{C_i} = [\boldsymbol{L}_{C_i}^{t-4}, \boldsymbol{L}_{C_i}^{t-3}, \boldsymbol{L}_{C_i}^{t-2}, \boldsymbol{L}_{C_i}^{t-1}, \boldsymbol{L}_{C_i}^{t}]
\end{equation}
where \( \boldsymbol{H}_{C_i} \) denotes the recorded local averages for mine \( C_i \) over the last five rounds, ensuring that short-term fluctuations are taken into account. Using this historical data, the average variance $V_{C_i}$ of the mine’s updates can be computed as: 
\begin{equation}
\boldsymbol{V}_{C_i} = \text{mean}(\text{var}(\boldsymbol{H}_{C_i}))
\end{equation}
where \( \boldsymbol{V}_{C_i} \) captures how much a mine's local averages fluctuate over recent rounds. Higher variance suggests erratic updates, a hallmark of additive noise attacks. To establish a {\bf detection threshold}, we compute the median and standard deviation of all \( \boldsymbol{V}_{C_i} \)  across all mines as:
\begin{equation}
\begin{split}
\boldsymbol{T}_v = \text{median}(\boldsymbol{V}_{C_1}, \boldsymbol{V}_{C_2}, \dots, \boldsymbol{V}_{C_N})\\
\quad + 2 \cdot \text{std}(\boldsymbol{V}_{C_1}, \boldsymbol{V}_{C_2}, \dots, \boldsymbol{V}_{C_N})
\end{split}
\end{equation}

This threshold \( \boldsymbol{T}_v \) ensures that only mines with significantly higher variance than the majority are flagged. 
\vspace{2mm}

{\noindent\bf B. Parameter Magnitude Analysis.} Another key characteristic of additive noise attacks is abnormally high parameter magnitudes due to injected noise. The magnitude of a mine’s local average is computed as:
\vspace{-0.3cm}
\begin{equation}
\boldsymbol{Z}_{C_i} = \| \boldsymbol{L}_{C_i}^t \|
\end{equation}
where \( \boldsymbol{Z}_{C_i} \) represents the norm of the local average at round \( t \) for mine \(C_i\). The {\bf global threshold} for parameter magnitudes is  then computed as:
\begin{equation}
\boldsymbol{T}_z = \boldsymbol{Z}_{\text{global}} + 2 \cdot \text{std}(\boldsymbol{Z}_{C_1}, ..., \boldsymbol{Z}_{C_N})
\end{equation}
\vspace{-0.1cm}
where \((\boldsymbol{Z}_{\text{global}} = \text{median}(\boldsymbol{Z}_{C_1}, \boldsymbol{Z}_{C_2}, \dots, \boldsymbol{Z}_{C_N})\) is the median of all \(\boldsymbol{Z}_{C_i})\), ensuring robustness against outliers. If a mine’s parameter magnitude is significantly larger than this threshold, it suggests possible additive noise attack.

\begin{algorithm}[!t]
\caption{Additive Noise Attack Detection Scheme}\label{algorithm3}
\KwIn{Each mine \( C_i \) where \( C_i \in \mathcal{C}_{B} \cup \mathcal{C}_{M} \cup \mathcal{C}_{U} \), Number of mines \( \mathcal{N} \)}
\KwOut{Set of additive noise attackers \( S_{\text{add}} \)}

$S_{\text{add}} \gets \emptyset$ \\

Compute local average \( \boldsymbol{L}_{C_i}^t \) for each mine \( C_i \) using Eq.~(6)\\
Compute global average \( \boldsymbol{G}^t \) using Eq.~(7)\\
Find \(\boldsymbol{H}_{C_i}\) using Eq.~(9)\\
Compute \(\boldsymbol{V}_{C_i}\) using Eq.~(10)\\
Compute \(\boldsymbol{T}_v\) using Eq.~(11)\\
Compute \(\boldsymbol{Z}_{C_i}\) using Eq.~(12)\\
\(Z_{\text{global}} = \text{median}(Z_{C_1}, Z_{C_2}, \dots, Z_{C_N})
\) \\

Compute \(\boldsymbol{T}_z \) using Eq.~(13)\\
 \For{$i \gets 1$ \textbf{to} $\mathcal{N}$}{
    \If{$\boldsymbol{V}_{C_i} > \boldsymbol{T}_v$ \textbf{or} $\boldsymbol{Z}_{C_i} > \boldsymbol{T}_z$}{
        $S_{\text{add}} \gets S_{\text{add}} \cup \{C_i\}$ \quad \text{// Detect additive noise attack} \\
    }
}

\Return $S_{\text{add}}$
\end{algorithm}
Algorithm~\ref{algorithm3} explains our structured detection scheme for additive noise attacks. 

\subsection{Unreliable Mines Identification Scheme}
Detecting unreliable mines is crucial, but instead of excluding them entirely, their impact on the global model should be reduced to prevent undue influence. To detect unreliable mines, Algorithm~\ref{algorithm4} calculates the Euclidean distance between each mine’s local update \( \boldsymbol{L}_{C_i}^t \) and the global average \( \boldsymbol{G}^t \). To simulate the effects of low-quality data, a small amount of Gaussian noise is introduced to the raw data before training commences. This can be mathematically expressed as:
\begin{equation}
\boldsymbol{X}_{C_i}' = \boldsymbol{X}_{C_i} + \mathcal{N}(0, \sigma^2)
\end{equation}
where \( \boldsymbol{X}_{C_i} \) denotes the original input data of mine \( C_i \),  \( \boldsymbol{X}_{C_i}' \) denotes the noisy version of the data used for training, and \( \mathcal{N}(0, \sigma^2) \) represents Gaussian noise drawn from a normal distribution with zero mean and variance \( \sigma^2 \). To effectively identify unreliable mines, the detection process relies on following two statistical measures:

\begin{itemize}[leftmargin=*]

    \item \textbf{Deviation Calculation}: The Euclidean distance between each mine’s local average and the global average is computed as:
    \begin{equation}
    \boldsymbol{E}_{C_i} = \| \boldsymbol{L}_{C_i}^t - \boldsymbol{G}^t \|
    \end{equation}
    where \( \boldsymbol{E}_{C_i} \) measures how far each mine's local average deviates from the central trend. Unlike simple thresholding techniques, Euclidean distance ensures that even small but consistent deviations are accounted for. 

    \item \textbf{Dynamic Threshold for Detection}: To detect outliers, a threshold is set using the mean and standard deviation of all deviations:
    \begin{equation}
   \Resize{7.5cm}{\boldsymbol{T}_E = \text{mean}(\boldsymbol{E}_{C_1}, \boldsymbol{E}_{C_2}, ..., \boldsymbol{E}_{C_N}) + \text{std}(\boldsymbol{E}_{C_1}, \boldsymbol{E}_{C_2}, ..., \boldsymbol{E}_{C_N})}
    \end{equation}
    The choice of a mean-based threshold ensures that the detection adapts to the training process dynamically. By incorporating the standard deviation, the threshold accounts for natural fluctuations in parameter updates, preventing false positives due to minor variations in mine updates. This approach is beneficial because it can adjust as training progresses.

\end{itemize}
Algorithm~\ref{algorithm4} describes the process MineDetect uses to identify unreliable mines.


\begin{algorithm}[!t]
\LinesNumbered 
\caption{Unreliable Mines Identification Scheme}\label{algorithm4}
\KwIn{Each mine \( C_i \) where \( C_i \in \mathcal{C}_{B} \cup \mathcal{C}_{M} \cup \mathcal{C}_{U} \), Number of mines \( \mathcal{N} \)}
\KwOut{Set of unreliable mines \( S_{\text{unrl}} \)}

$S_{\text{unrl}} \gets \emptyset$ \\
Compute local average \( \boldsymbol{L}_{C_i}^t \) for each mine \( C_i \) using Eq.~(6)\\
Compute global average \( \boldsymbol{G}^t \) using Eq.~(7)\\

Compute \(\boldsymbol{E}_{C_i}\) using Eq.~(15)\\

Compute \(\boldsymbol{T}_E\) using Eq.~(16) \tcp{ Threshold}

\For{$i \gets 1$ \textbf{to} $\mathcal{N}$}{ 
    \If{$\boldsymbol{E}_{C_i} > \boldsymbol{T}_E$}{ 
        $S_{\text{unrl}} \gets S_{\text{unrl}} \cup \{C_i\}$ \tcp{Mark Mine as Unreliable}
    }
}

\Return $S_{\text{unrl}}$ 
\end{algorithm}

\section{Complexity Analysis of MineDetect Algorithms}

The MineDetect framework consists of three detection algorithms: {\em sign-flip attack detection} (Algorithm~\ref{algorithm2}), {\em additive noise attack detection} (Algorithm~\ref{algorithm3}), and {\em unreliable mine Identification} (Algorithm~\ref{algorithm4}). Let \( p \) represent the number of model parameters in a local model. The {\em cosine similarity} calculation for each mine takes \( \mathcal{O}(p) \). The {\em sign-flip attack detection} algorithm computes the local and global averages in \( \mathcal{O}(N) \), resulting in a total time complexity of \( \mathcal{O}(Np) \). For {\em additive noise attack detection}, the algorithm computes local and global averages in \( \mathcal{O}(N) \), calculates historical variance over five rounds in \( \mathcal{O}(N) \), and evaluates thresholds in \( \mathcal{O}(N) \) making total time complexity as \( \mathcal{O}(Np) \). The {\em unreliable mine detection} algorithm computes Euclidean distances between each mine's local average and the global average in \( \mathcal{O}(Np) \), determines the detection threshold in \( \mathcal{O}(N) \), and classifies the mines in \( \mathcal{O}(N) \) making overall time complexity as \( \mathcal{O}(Np) \). Since the three algorithms are executed sequentially, the overall per-round time complexity of the MineDetect framework remains \( \mathcal{O}(Np) \), ensuring scalability. Regarding space complexity, MineDetect requires \( \mathcal{O}((N+1)p) \) space to store \( N \) local averages (one per mine) and a single global average for each round. This is significantly more efficient compared to approaches like the one in \cite{gupta2022long}, where storing 3 previous parameters for each mine and one summation of parameters results in a total  of \( \mathcal{O}(4Np) \). By storing only the essential statistics and threshold values, MineDetect ensures memory efficiency while maintaining robustness.

\begin{table}[!t]
\caption{Simulation Parameters.}
\label{parameters-table}
\centering
\renewcommand{\arraystretch}{1} 
\begin{tabular}{c|ccc}
\toprule
                                          & \multicolumn{3}{c}{\textbf{Dataset}}                                                                                     \\ \cline{2-4} 
\multirow{-2}{*}{\textbf{Hyperparameter}} & \multicolumn{1}{c|}{\textbf{MNIST}} & \multicolumn{1}{c|}{\textbf{Fashion-MNIST}} & \textbf{Rockburst} \\ \hline
Number of mines                            & \multicolumn{1}{c|}{40}   & \multicolumn{1}{c|}{40}            & 40        \\ \hline

Learning Rate                             & \multicolumn{1}{c|}{0.01}  & \multicolumn{1}{c|}{0.01}          & 0.01      \\ \hline
Batch Size                                & \multicolumn{1}{c|}{64}    & \multicolumn{1}{c|}{64}            & 8         \\ \hline
Optimizer                                 & \multicolumn{1}{c|}{SGD}   & \multicolumn{1}{c|}{SGD}           & SGD       \\ \hline
Local Epochs                              & \multicolumn{1}{c|}{4}     & \multicolumn{1}{c|}{4}             & 60        \\ \hline
Weight Decay                              & \multicolumn{1}{c|}{0.0001} & \multicolumn{1}{c|}{0.0001}        & 0.0001    \\ \hline
Momentum                                  & \multicolumn{1}{c|}{0.9}  & \multicolumn{1}{c|}{0.9}           & 0.9       \\ \bottomrule 
\end{tabular}
\end{table}
\section{Performance Evaluation}
\subsection{Experimental Setup}
{\bf Datasets.} To assess and demonstrate the generalizability our MineDetect framework against state-of-the-art approaches, We have considered a couple of benchmark datasets like MNIST \cite{deng2012mnist}, Fashion-MNIST \cite{xiao2017fashion} along with Rockburst classification Dataset taken from the authors of \cite{shirani2024hybridized}. This dataset is designed for classifying {\em rockburst risk levels in underground mines} categorized into {\bf four} risk levels: none, light, moderate, and severe. Due to its limited size, we have applied the Synthetic Minority Over-Sampling Technique (SMOTE)~\cite{chawla2002smote}, expanding the dataset to 1,000 samples.

{\noindent \bf Local  Models.} We have employed the same local models from \cite{gupta2022long} for the MNIST and Fashion-MNIST datasets. For the customized rockburst classification dataset, we have created following  lightweight neural network model: 
\begin{itemize}[leftmargin=*]
    \item Seven input neurons, each representing one of the seven features used to predict rockburst risk levels.
    \item Three hidden layers with 128, 64, and 32 neurons, respectively, using ReLU activation functions with a dropout rate of 0.4 after the first two hidden layers to prevent overfitting.
    \item Four output neurons, each corresponding to one of the four rockburst risk levels.
\end{itemize}
The hyperparameters for training the local models, are summarized in Table \ref{parameters-table}.


{\noindent \bf Environment.} We have conducted our simulations on Google Colaboratory \cite{bisong2019google}, utilizing an NVIDIA T4 GPU, 50.99 GB of RAM, and 235.68 GB of storage. Various versions of Python 3 and packages from the PyTorch framework are also installed.

{\noindent \bf Evaluation Metrics.} To thoroughly evaluate the effectiveness of our MineDetect framework, we have used following evaluation metrics:

\begin{itemize}[leftmargin=*]
\item {\bf Minimum Accuracy:} It is the minimum accuracy value observed among all global rounds $T$.
\item {\bf Maximum Accuracy:} It represents the highest accuracy achieved across all global rounds $T$.
\item {\bf Average Accuracy:} It is the mean accuracy across all global rounds $T$.
\item {\bf Confusion Matrix:} It summarizes classification results with four components: TN (true negatives), TP (true positives), FP (negatives misclassified as positives), and FN (positives misclassified as negatives). 


\item {\bf False Positive Rate (FPR)}: FPR is the proportion of negatives misclassified as positives, given by $\text{FPR} = \text{FP}/(\text{FP} + \text{TN})$. A lower FPR reflects better class distinction.
\end{itemize}

\noindent{\bf Baselines.}
 we have evaluated MineDetect's detection performance against several FL defense mechanisms, including Krum~\cite{blanchard2017machine}, Mkrum~\cite{blanchard2017machine}, GeoMed~\cite{chen2017distributed}, and MUD-HoG~\cite{gupta2022long}.

\begin{table}[!]
\caption{Comparison of MineDetect vs. baselines in terms of Min, Max, and Avg accuracy for the RockBurst Classification Dataset.}
\label{tab:comp-accu-rock}
\centering
\resizebox{\columnwidth}{!}{%
\begin{tabular}{c|ccccc}
\toprule
& \multicolumn{5}{c}{\textbf{Accuracy (\%)}}                                                                                              \\ \cline{2-6} \multirow{-2}{*}{\textbf{Round}}                                                              & \multicolumn{1}{c|}{krum}                                                   & \multicolumn{1}{c|}{mkrum}                                                                               & \multicolumn{1}{c|}{GeoMed}                                                                              & \multicolumn{1}{c|}{Mud-HoG}                                                                             & {\bf MineDetect  }                                                                                 \\ \hline
1                                                                                             & \multicolumn{1}{c|}{{\color[HTML]{1F1F1F} 25}}                              & \multicolumn{1}{c|}{{\color[HTML]{1F1F1F} 54.0}}                                                         & \multicolumn{1}{c|}{{\color[HTML]{1F1F1F} 58.0}}                                                         & \multicolumn{1}{c|}{{\color[HTML]{1F1F1F} 42.5}}                                                         & {\color[HTML]{1F1F1F} 57.5}                                                                  \\ \hline
2                                                                                             & \multicolumn{1}{c|}{25}                                                     & \multicolumn{1}{c|}{{\color[HTML]{1F1F1F} 53.5}}                                                         & \multicolumn{1}{c|}{{\color[HTML]{1F1F1F} 57.5}}                                                         & \multicolumn{1}{c|}{{\color[HTML]{1F1F1F} 60.0}}                                                         & {\color[HTML]{1F1F1F} 59.0}                                                                  \\ \hline
3                                                                                             & \multicolumn{1}{c|}{25}                                                     & \multicolumn{1}{c|}{{\color[HTML]{1F1F1F} 56.5}}                                                         & \multicolumn{1}{c|}{{\color[HTML]{1F1F1F} 58.5}}                                                         & \multicolumn{1}{c|}{{\color[HTML]{1F1F1F} 62.0}}                                                         & {\color[HTML]{1F1F1F} 62.0}                                                                  \\ \hline
4                                                                                             & \multicolumn{1}{c|}{{\color[HTML]{1F1F1F} 25}}                              & \multicolumn{1}{c|}{{\color[HTML]{1F1F1F} 58.5}}                                                         & \multicolumn{1}{c|}{{\color[HTML]{1F1F1F} 58.5}}                                                         & \multicolumn{1}{c|}{{\color[HTML]{1F1F1F} 61.0}}                                                         & {\color[HTML]{1F1F1F} 60.5}                                                                  \\ \hline
5                                                                                             & \multicolumn{1}{c|}{{\color[HTML]{1F1F1F} 25}}                              & \multicolumn{1}{c|}{{\color[HTML]{1F1F1F} 60.5}}                                                         & \multicolumn{1}{c|}{{\color[HTML]{1F1F1F} 59.5}}                                                         & \multicolumn{1}{c|}{{\color[HTML]{1F1F1F} 63.5}}                                                         & {\color[HTML]{1F1F1F} 62.0}                                                                  \\ \hline
6                                                                                             & \multicolumn{1}{c|}{{\color[HTML]{1F1F1F} 25}}                            & \multicolumn{1}{c|}{{\color[HTML]{1F1F1F} 62.0}}                                                         & \multicolumn{1}{c|}{{\color[HTML]{1F1F1F} 60.5}}                                                         & \multicolumn{1}{c|}{{\color[HTML]{1F1F1F} 63.0}}                                                         & {\color[HTML]{1F1F1F} 62.5}                                                                  \\ \hline
7                                                                                             & \multicolumn{1}{c|}{{\color[HTML]{1F1F1F} 25}}                              & \multicolumn{1}{c|}{{\color[HTML]{1F1F1F} 60.0}}                                                         & \multicolumn{1}{c|}{{\color[HTML]{1F1F1F} 60.5}}                                                         & \multicolumn{1}{c|}{{\color[HTML]{1F1F1F} 64.0}}                                                         & {\color[HTML]{1F1F1F} 63.5}                                                                  \\ \hline
8                                                                                             & \multicolumn{1}{c|}{{\color[HTML]{1F1F1F} 25}}                              & \multicolumn{1}{c|}{{\color[HTML]{1F1F1F} 60.5}}                                                         & \multicolumn{1}{c|}{{\color[HTML]{1F1F1F} 62.0}}                                                         & \multicolumn{1}{c|}{{\color[HTML]{1F1F1F} 63.5}}                                                         & {\color[HTML]{1F1F1F} 67.0}                                                                  \\ \hline
9                                                                                             & \multicolumn{1}{c|}{{\color[HTML]{1F1F1F} 25}}                              & \multicolumn{1}{c|}{{\color[HTML]{1F1F1F} 60.5}}                                                         & \multicolumn{1}{c|}{{\color[HTML]{1F1F1F} 62.0}}                                                         & \multicolumn{1}{c|}{{\color[HTML]{1F1F1F} 64.0}}                                                         & {\color[HTML]{1F1F1F} 66.5}                                                                  \\ \hline
10                                                                                            & \multicolumn{1}{c|}{{\color[HTML]{1F1F1F} 25}}                              & \multicolumn{1}{c|}{{\color[HTML]{1F1F1F} 61.0}}                                                         & \multicolumn{1}{c|}{{\color[HTML]{1F1F1F} 62.5}}                                                         & \multicolumn{1}{c|}{{\color[HTML]{1F1F1F} 65.5}}                                                         & {\color[HTML]{1F1F1F} 66.5}                                                                  \\ \hline
11                                                                                            & \multicolumn{1}{c|}{25}                                                     & \multicolumn{1}{c|}{{\color[HTML]{1F1F1F} 60.5}}                                                         & \multicolumn{1}{c|}{{\color[HTML]{1F1F1F} 62.0}}                                                         & \multicolumn{1}{c|}{{\color[HTML]{1F1F1F} 66.0}}                                                         & {\color[HTML]{1F1F1F} 68.5}                                                                  \\ \hline
12                                                                                            & \multicolumn{1}{c|}{25}                                                     & \multicolumn{1}{c|}{{\color[HTML]{1F1F1F} 62.5}}                                                         & \multicolumn{1}{c|}{{\color[HTML]{1F1F1F} 62.5}}                                                         & \multicolumn{1}{c|}{{\color[HTML]{1F1F1F} 65.5}}                                                         & {\color[HTML]{1F1F1F} 69.5}                                                                  \\ \hline
13                                                                                            & \multicolumn{1}{c|}{25}                                                     & \multicolumn{1}{c|}{{\color[HTML]{1F1F1F} 61.5}}                                                         & \multicolumn{1}{c|}{{\color[HTML]{1F1F1F} 64.0}}                                                         & \multicolumn{1}{c|}{{\color[HTML]{1F1F1F} 66.5}}                                                         & {\color[HTML]{1F1F1F} 69.0}                                                                  \\ \hline
14                                                                                            & \multicolumn{1}{c|}{25}                                                     & \multicolumn{1}{c|}{{\color[HTML]{1F1F1F} 62.0}}                                                         & \multicolumn{1}{c|}{{\color[HTML]{1F1F1F} 64.5}}                                                         & \multicolumn{1}{c|}{{\color[HTML]{1F1F1F} 66.0}}                                                         & {\color[HTML]{1F1F1F} 67.5}                                                                  \\ \hline
15                                                                                            & \multicolumn{1}{c|}{25}                                                     & \multicolumn{1}{c|}{{\color[HTML]{1F1F1F} 60.5}}                                                         & \multicolumn{1}{c|}{{\color[HTML]{1F1F1F} 67.5}}                                                         & \multicolumn{1}{c|}{{\color[HTML]{1F1F1F} 66.0}}                                                         & {\color[HTML]{1F1F1F} 69.0}                                                                  \\ \hline
16                                                                                            & \multicolumn{1}{c|}{25}                                                     & \multicolumn{1}{c|}{{\color[HTML]{1F1F1F} 60.5}}                                                         & \multicolumn{1}{c|}{{\color[HTML]{1F1F1F} 67.0}}                                                         & \multicolumn{1}{c|}{{\color[HTML]{1F1F1F} 68.0}}                                                         & {\color[HTML]{1F1F1F} 70.0}                                                                  \\ \hline
17                                                                                            & \multicolumn{1}{c|}{25}                                                     & \multicolumn{1}{c|}{{\color[HTML]{1F1F1F} 61.0}}                                                         & \multicolumn{1}{c|}{{\color[HTML]{1F1F1F} 69.0}}                                                         & \multicolumn{1}{c|}{{\color[HTML]{1F1F1F} 69.0}}                                                         & {\color[HTML]{1F1F1F} 72.0}                                                                  \\ \hline
18                                                                                            & \multicolumn{1}{c|}{25}                                                     & \multicolumn{1}{c|}{{\color[HTML]{1F1F1F} 61.5}}                                                         & \multicolumn{1}{c|}{{\color[HTML]{1F1F1F} 67.0}}                                                         & \multicolumn{1}{c|}{{\color[HTML]{1F1F1F} 70.0}}                                                         & {\color[HTML]{1F1F1F} 71.0}                                                                  \\ \hline
19                                                                                            & \multicolumn{1}{c|}{25}                                                     & \multicolumn{1}{c|}{{\color[HTML]{1F1F1F} 62.5}}                                                         & \multicolumn{1}{c|}{{\color[HTML]{1F1F1F} 68.5}}                                                         & \multicolumn{1}{c|}{{\color[HTML]{1F1F1F} 71.5}}                                                         & {\color[HTML]{1F1F1F} 72.5}                                                                  \\ \hline
20                                                                                            & \multicolumn{1}{c|}{25}                                                     & \multicolumn{1}{c|}{{\color[HTML]{1F1F1F} 62.5}}                                                         & \multicolumn{1}{c|}{{\color[HTML]{1F1F1F} 68.0}}                                                         & \multicolumn{1}{c|}{{\color[HTML]{1F1F1F} 71.5}}                                                         & {\color[HTML]{1F1F1F} 73.5}                                                                  \\ 
\midrule
\textbf{\begin{tabular}[c]{@{}c@{}}Min Accuracy \\ Max Accuracy \\ Avg Accuracy\end{tabular}} & \multicolumn{1}{c|}{\begin{tabular}[c]{@{}c@{}}25 \\ 25 \\ 25\end{tabular}} & \multicolumn{1}{c|}{{\color[HTML]{1F1F1F} \begin{tabular}[c]{@{}c@{}}53.5 \\ 62.5 \\ 60.1\end{tabular}}} & \multicolumn{1}{c|}{{\color[HTML]{1F1F1F} \begin{tabular}[c]{@{}c@{}}57.5 \\ 69.0 \\ 63.0\end{tabular}}} & \multicolumn{1}{c|}{{\color[HTML]{1F1F1F} \begin{tabular}[c]{@{}c@{}}42.5 \\ 71.5 \\ 64.4\end{tabular}}} & {\color[HTML]{1F1F1F} \textbf{\begin{tabular}[c]{@{}c@{}}57.5 \\ 73.5 \\ 66.5\end{tabular}}} \\ \bottomrule
\end{tabular}
}\vspace{-2mm}

\end{table}

\begin{table}[!t]
\caption{Comparison of MineDetect vs. baselines in terms of Min, Max, and Avg accuracy for the MNIST Dataset.}
 
\label{tab:comp-accu-mnist}
\centering
\resizebox{\columnwidth}{!}{%
\begin{tabular}{c|ccccc}
\toprule
\multirow{2}{*}{Round} & \multicolumn{5}{c}{Accuracy (\%)}                                                                                                             \\ \cline{2-6} 
                       & \multicolumn{1}{c|}{krum}  & \multicolumn{1}{c|}{mkrum}          & \multicolumn{1}{c|}{GeoMed} & \multicolumn{1}{c|}{MUD-HoG} & {\bf MineDetect}     \\ \hline
1                      & \multicolumn{1}{c|}{58.66} & \multicolumn{1}{c|}{75.58}          & \multicolumn{1}{c|}{71.35}  & \multicolumn{1}{c|}{64.1}    & 77.26          \\ \hline
2                      & \multicolumn{1}{c|}{74.17} & \multicolumn{1}{c|}{87.36}          & \multicolumn{1}{c|}{88.08}  & \multicolumn{1}{c|}{76.79}   & 68.48          \\ \hline
3                      & \multicolumn{1}{c|}{78.16} & \multicolumn{1}{c|}{92.99}          & \multicolumn{1}{c|}{92.68}  & \multicolumn{1}{c|}{90.93}   & 93.24          \\ \hline
4                      & \multicolumn{1}{c|}{89.67} & \multicolumn{1}{c|}{94.73}          & \multicolumn{1}{c|}{94.44}  & \multicolumn{1}{c|}{87.75}   & 90.71          \\ \hline
5                      & \multicolumn{1}{c|}{89.43} & \multicolumn{1}{c|}{95.57}          & \multicolumn{1}{c|}{95.66}  & \multicolumn{1}{c|}{95.4}    & 96.19          \\ \hline
6                      & \multicolumn{1}{c|}{86.89} & \multicolumn{1}{c|}{96.25}          & \multicolumn{1}{c|}{96.18}  & \multicolumn{1}{c|}{95.98}   & 95.56          \\ \hline
7                      & \multicolumn{1}{c|}{91.05} & \multicolumn{1}{c|}{96.45}          & \multicolumn{1}{c|}{96.61}  & \multicolumn{1}{c|}{96.33}   & 96.54          \\ \hline
8                      & \multicolumn{1}{c|}{91.55} & \multicolumn{1}{c|}{96.87}          & \multicolumn{1}{c|}{96.72}  & \multicolumn{1}{c|}{96.83}   & 96.72          \\ \hline
9                      & \multicolumn{1}{c|}{91.31} & \multicolumn{1}{c|}{96.99}          & \multicolumn{1}{c|}{97.15}  & \multicolumn{1}{c|}{96.99}   & 97.1           \\ \hline
10                     & \multicolumn{1}{c|}{91.44} & \multicolumn{1}{c|}{97.23}          & \multicolumn{1}{c|}{97.13}  & \multicolumn{1}{c|}{97.05}   & 97.27          \\ 

\midrule
\textbf{Min Accuracy}           & \multicolumn{1}{c|}{58.66} & \multicolumn{1}{c|}{\textbf{75.58}} & \multicolumn{1}{c|}{71.35}  & \multicolumn{1}{c|}{64.1}    & 68.48          \\ 
\textbf{Max Accuracy}           & \multicolumn{1}{c|}{91.55} & \multicolumn{1}{c|}{97.23}          & \multicolumn{1}{c|}{97.15}  & \multicolumn{1}{c|}{97.05}   & \textbf{97.27} \\
\textbf{Avg Accuracy}           & \multicolumn{1}{c|}{84.24} & \multicolumn{1}{c|}{\textbf{93.03}} & \multicolumn{1}{c|}{92.79}  & \multicolumn{1}{c|}{89.91}   & 90.91          \\ \bottomrule
\end{tabular}
}

\end{table}

\begin{table}[!t]
 
\caption{Comparison of MineDetect vs. baselines in terms of Min, Max, and Avg accuracy for the Fashion-MNIST Dataset.}
\label{tab:comp-accu-fmnist}
\centering
\resizebox{\columnwidth}{!}{%
\begin{tabular}{c|c|c|c|c|c}
\toprule
 & \multicolumn{5}{c}{\textbf{Accuracy (\%)}}   \\ \cline{2-6} 
\multirow{-2}{*}{\textbf{Round}} & \multicolumn{1}{c|}{krum}                         & \multicolumn{1}{c|}{mkrum}                        & \multicolumn{1}{c|}{GeoMed}                                & \multicolumn{1}{c|}{MUD-HoG}                               & {\bf MineDetect }                  \\ \hline
1                                & 10.14 & 10.0  & 10.0           & 15.11          & 10.01 \\ \hline
2                                & 10.0  & 14.70 & 61.95          & 64.15          & 68.21 \\ \hline
3                                & 10.0  & 71.39 & 72.6           & 61.71          & 49.88 \\ \hline
4                                & 10.0  & 74.99 & 73.64          & 63.25          & 74.69 \\ \hline
5                                & 10.0  & 76.76 & 76.18          & 77.02          & 72.26 \\ \hline
6                                & 10.0  & 78.43 & 77.97          & 79.65          & 79.75 \\ \hline
7                                & 10.0  & 79.68 & 79.73          & 81.83          & 79.2  \\ \hline
8                                & 10.0  & 80.77 & 81.70          & 83.5           & 79.18 \\ \hline
9                                & 10.0  & 81.68 & 83.24          & 84.1           & 81.93 \\ \hline
10                               & 10.0  & 82.19 & 83.92          & 84.1           & 82.18 \\ \midrule

\textbf{Min Accuracy}            & 10                           & 10.0  & 10.0           & \textbf{15.11} & 10.01 \\ 
\textbf{Max Accuracy}            & 10.14                        & 82.19 & 83.92          & \textbf{84.1}  & 82.18 \\ 
\textbf{Avg Accuracy}            & 10.014                       & 67.36 & \textbf{72.12} & 71.96          & 67.93 \\ \bottomrule
\end{tabular}
}
\end{table}

\begin{figure*}[!t]
\centering
    \begin{subfigure}[b]{0.32\textwidth}
        \centering
        \includegraphics[width=\linewidth, height=5cm]{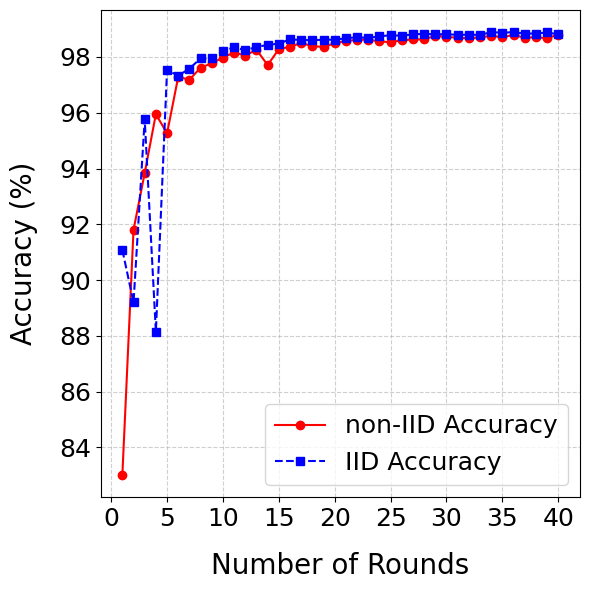}
        \caption{MNIST Dataset}
        \label{fig:mnistiid-noniid}
    \end{subfigure}
    \begin{subfigure}[b]{0.32\textwidth}
        \centering
        \includegraphics[width=\linewidth, height=5cm]{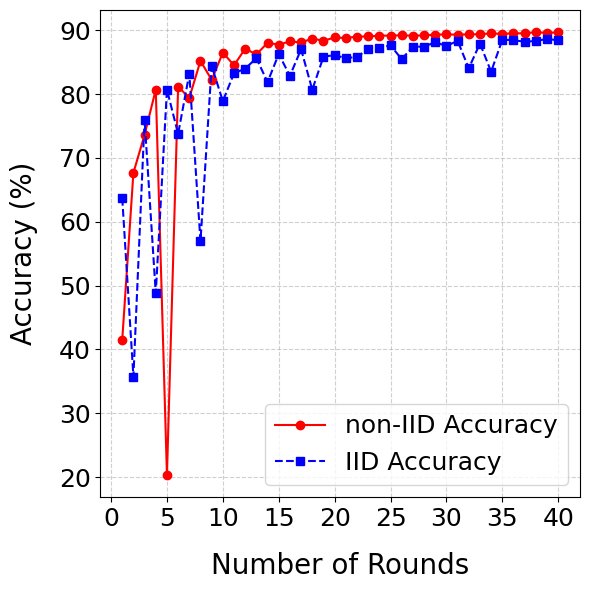}
        \caption{Fashion-MNIST Dataset}
        \label{fig:fmnistiid-noniid}
    \end{subfigure}
    \begin{subfigure}[b]{0.32\textwidth}
        \centering
        \includegraphics[width=\linewidth, height=5cm]{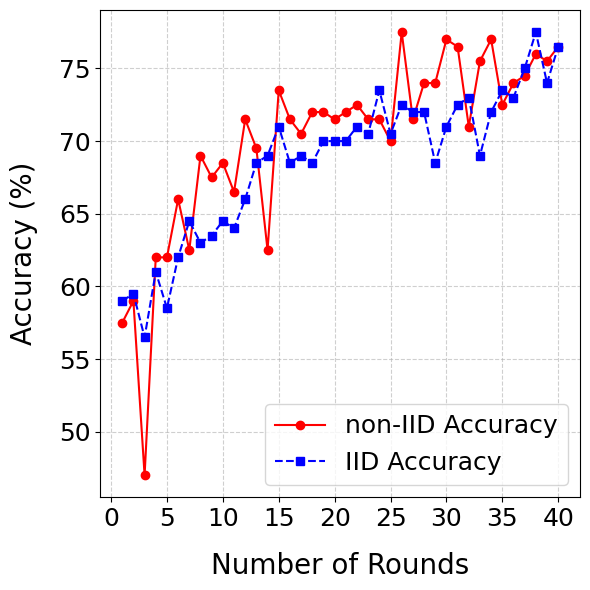}
        \caption{Rockburst Classification Dataset}
        \label{fig:rockiid-noniid}
    \end{subfigure}
  \caption{MineDetect's accuracy under both IID and Non-IID data distributions for (a) MNIST, (b) Fashion-MNIST, and (c) Rockburst classification datasets.}

    \label{fig:iid-noniid-comparison}
\end{figure*}

\begin{figure*}[!]
    \centering
    \begin{subfigure}[b]{0.32\textwidth}
        \centering
        \includegraphics[width=\linewidth, height=5cm]{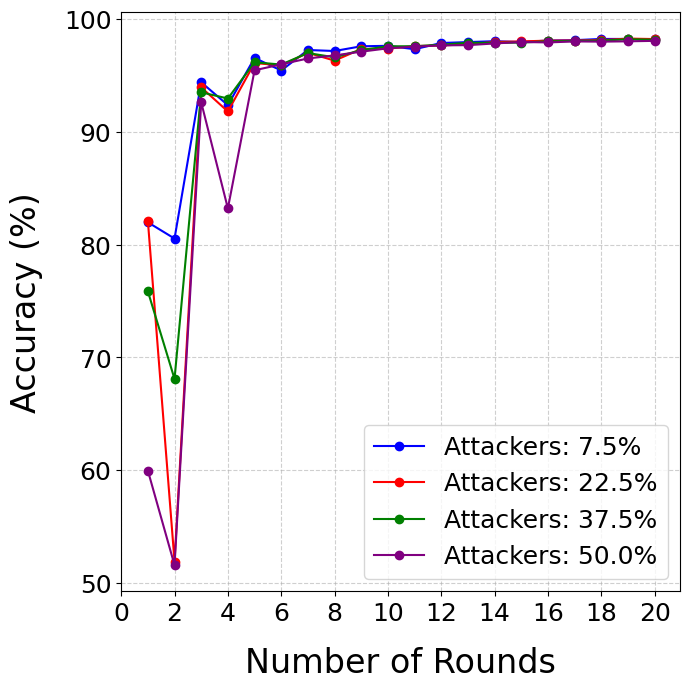}
        \caption{MNIST Dataset.}
        \label{fig:attackervs-accuracy-mnist}
    \end{subfigure}
    \begin{subfigure}[b]{0.32\textwidth}
        \centering
        \includegraphics[width=\linewidth, height=5cm]{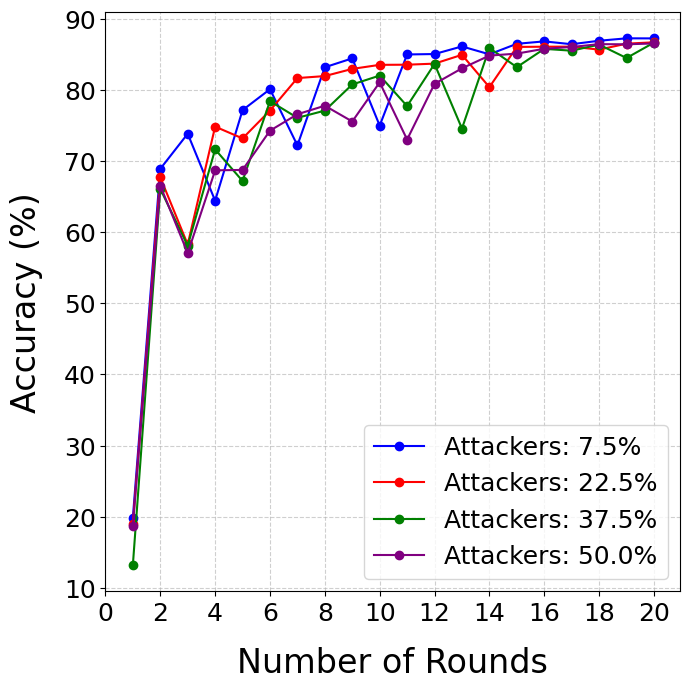}
        \caption{Fashion MNIST Dataset.}
        \label{fig:attackervs-accuracy-fmnist}
    \end{subfigure}
    \begin{subfigure}[b]{0.32\textwidth}
        \centering
        \includegraphics[width=\linewidth, height=5cm]{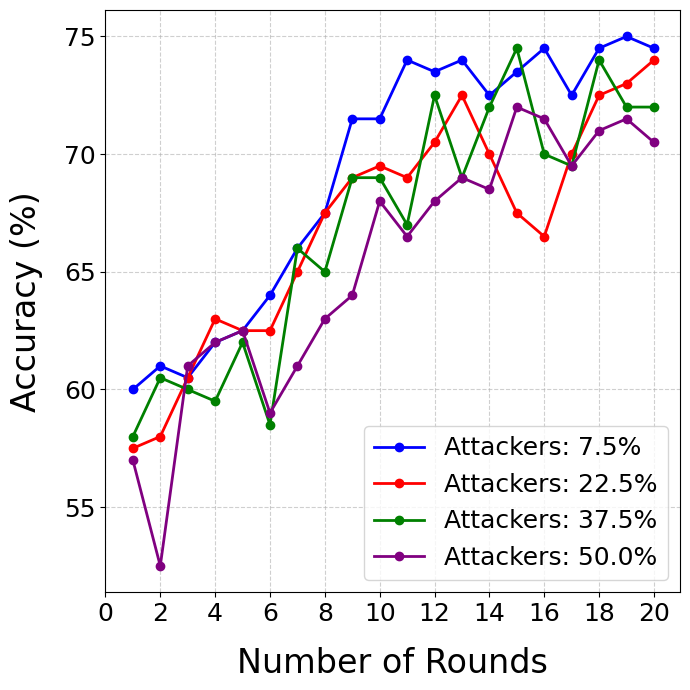}
        \caption{Rockburst Classification Dataset.}
        \label{fig:attackervs-accuracy-rock}
    \end{subfigure}
    
\caption{MineDetect's accuracy for varying attacker percentages (7.5\%, 22.5\%,37.5\%, and 50\%) for (a) MNIST, (b) Fashion-MNIST, and (c) Rockburst classification datasets.}\vspace{-2mm}

    \label{fig:attackervs-accuracy-all}
\end{figure*}

\subsection{Experimental Results}

{\em  {\bf 6.2.1 Comparing MineDetect vs. Baselines}}

\noindent MineDetect's performance is evaluated against the baselines using the Rockburst, MNIST, and Fashion-MNIST datasets. The evaluation considers different numbers of global rounds and local epochs based on model complexity: Rockburst (20 global rounds, 60 local epochs), MNIST (10 global rounds, 4 local epochs), and Fashion-MNIST (10 global rounds, 4 local epochs), all under a {\em non-IID distribution with 12 attackers among 40}. Among all techniques, MineDetect achieves the highest maximum accuracy of 73.5\% and the best average accuracy of 66. 5\% using the Rockburst classification dataset, real-world mining dataset, in 20 global rounds (see Table~\ref{tab:comp-accu-rock}). Furthermore, MineDetect demonstrates its capability to handle simpler datasets with heavy local models, achieving the highest maximum accuracy of 97.27\% and a strong average accuracy of 90.91\% on the MNIST dataset (see Table~\ref{tab:comp-accu-mnist}). On the Fashion-MNIST dataset (see Table~\ref{tab:comp-accu-fmnist}), MineDetect attains a maximum accuracy of 82.18\% and an average accuracy of 67.93\%, showcasing competitive performance. Despite the challenging non-IID distribution, it exhibits steady improvements over iterations, closely aligned with MUD-HoG, which achieves a slightly higher average accuracy of 71.96\%. Overall, MineDetect consistently outperforms or remains on par with state-of-the-art methodologies across all datasets. Its ability to maintain high accuracy, particularly in non-IID settings, underscores its robustness.

\vspace{2mm}
\noindent{\em {\bf 6.2.2 In-depth Evaluation of MineDetect}}

\noindent {\bf Effect of Data Distribution.} We have generated non-IID data for each dataset using a {\em Dirichlet distribution} to allocate class samples unevenly among mines. First, each class is identified and then randomly assigned to mines using Dirichlet probabilities controlled by the $\lambda$ parameter. A lower value of $\lambda$ results in highly skewed distributions, while a higher $\lambda$ leads to a more balanced (IID-like) allocation. We set $\lambda = 0.9$, which means each client likely has a mix of several classes rather than being dominated by a single class. 
In \fref{fig:iid-noniid-comparison}, we show the MineDetect framework's performance under IID and non-IID distributions for the three datasets across 40 global rounds. With both distributions converging quickly, ~\fref{fig:mnistiid-noniid} (MNIST) shows that simpler datasets are less impacted by data heterogeneity. Notable accuracy differences in the early rounds for the non-IID distribution in \fref{fig:fmnistiid-noniid} (Fashion-MNIST) highlight the destabilizing effect of diverse data distribution. Compared to IID, \fref{fig:rockiid-noniid} (Rockburst) similarly shows less convergence and more variability under non-IID conditions. MineDetect is resolute, steadily narrowing the accuracy disparity and achieving competitive performance—particularly evident in \fref{fig:fmnistiid-noniid} and \fref{fig:rockiid-noniid}, where data imbalance is more pronounced. These comparisons emphasize the need for resilient aggregation techniques such as MineDetect to maintain stability and precision among non-uniform data distributions.

\vspace{1mm}
\noindent{\bf {Effect of Attacker Percentage on Accuracy.}}  \fref{fig:attackervs-accuracy-mnist} (MNIST), \fref{fig:attackervs-accuracy-fmnist} (Fashion-MNIST), and \fref{fig:attackervs-accuracy-rock} (Rockburst) illustrate MineDetect's performance under varying percentages of attackers, revealing a clear inverse relationship between the proportion of attackers and model accuracy. The highest accuracy is consistently observed when only 7.5\% of mines are attackers, while accuracy decreases as the attacker percentage increases.  
When 50\% of the mines are attackers, the model experiences a significant initial accuracy drop but gradually recovers over subsequent rounds. However, even with this recovery, accuracy remains lower than in scenarios with fewer attackers throughout the training process. This trend demonstrates MineDetect’s resilience against adversarial behavior while also highlighting its sensitivity to high proportions of malicious mines.


\begin{figure*}[!t]
    \centering
    \begin{subfigure}{0.32\textwidth}
        \centering
        \includegraphics[width=\linewidth]{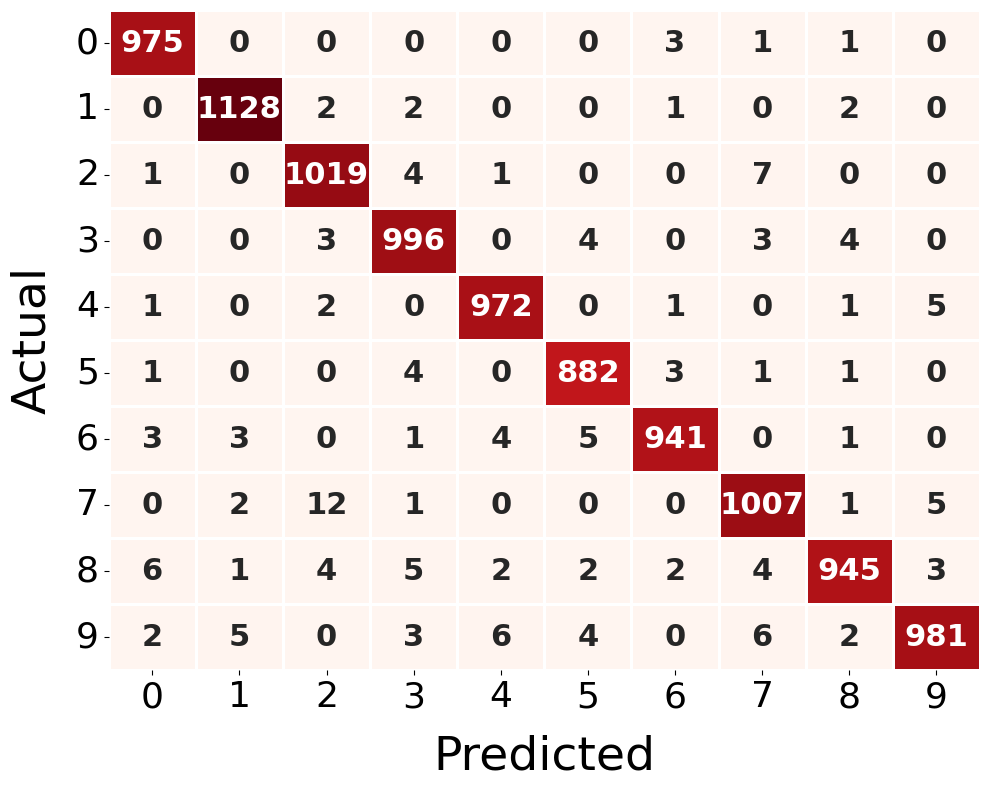}
        \caption{MNIST Dataset.}
        \label{fig:mnistcf}
    \end{subfigure}
    \begin{subfigure}{0.32\textwidth}
        \centering
    \includegraphics[width=\linewidth]{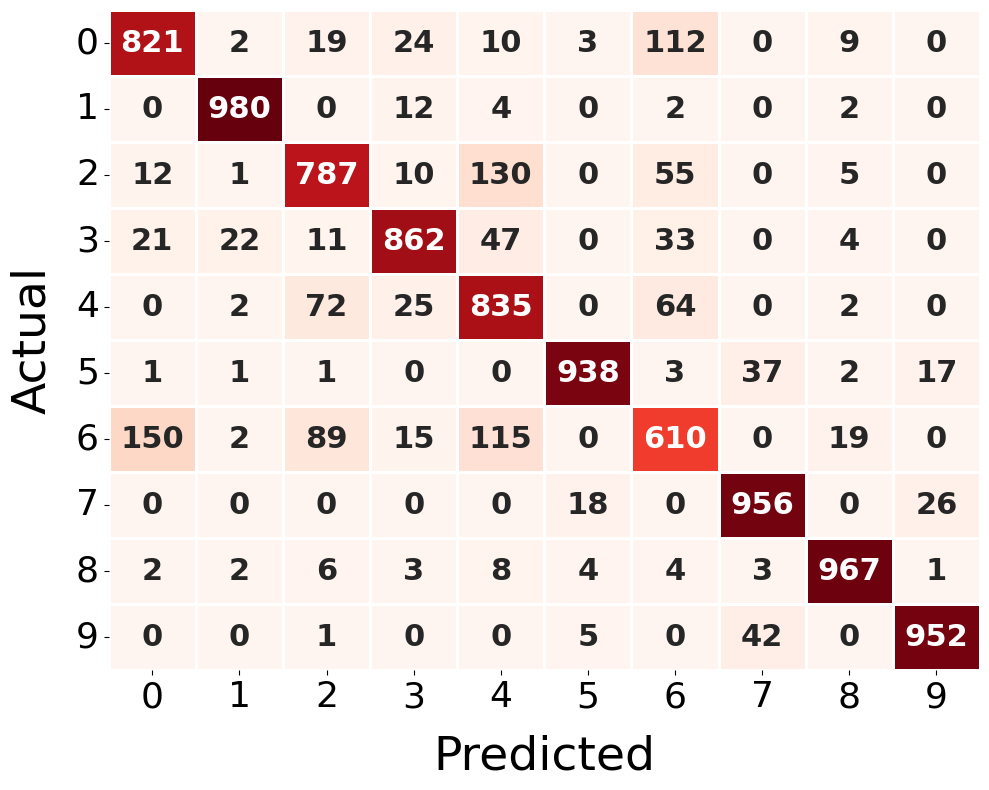}
        \caption{Fashion MNIST Dataset.}
        \label{fig:fmnistcf}
    \end{subfigure}
    \begin{subfigure}{0.33\textwidth}
        \centering
    \includegraphics[width=\linewidth]{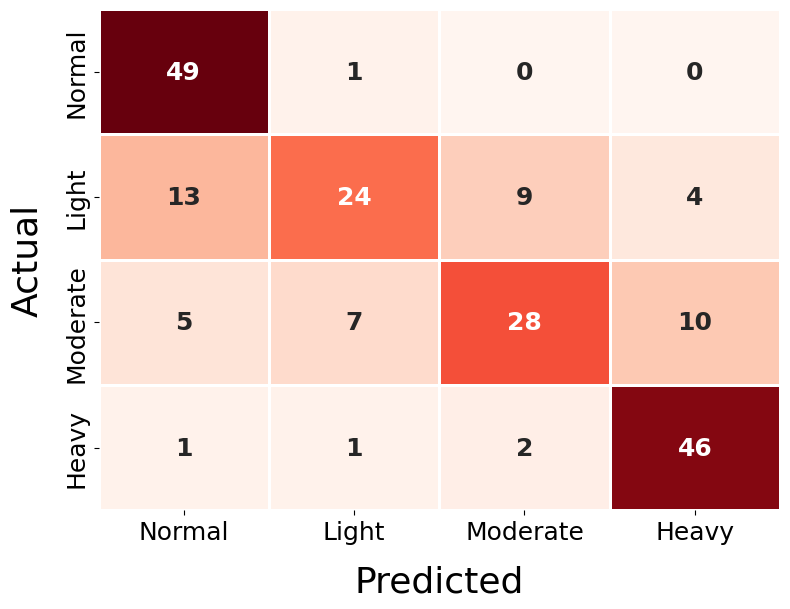}
        \caption{Rockburst Classification Dataset.}
        \label{fig:rockcf}
    \end{subfigure}
    \caption{Results of MineDetect's Confusion Matrix for (a) MNIST, (b) Fashion-MNIST, and (c) Rockburst classification Datasets.}
    \label{fig:all_cfs}\vspace{-2mm}
\end{figure*}

\vspace{2mm}
\noindent{\bf Evaluating MineDetect using a Confusion Matrix.}  Training in 40 global rounds, ~\fref{fig:mnistcf} (MNIST), ~\fref{fig:fmnistcf} (Fashion-MNIST), and ~\fref{fig:rockcf} (Rockburst) show the confusion matrix of the MineDetect framework. Four local epochs are used for the more complicated MNIST and Fashion-MNIST models, but the lightweight Rockburst model is trained for sixty local epochs to increase performance. Few misclassifications in ~\fref{fig:mnistcf} (MNIST) indicate exceptional pattern recognition. Similarly, ~\fref{fig:fmnistcf} (Fashion-MNIST) shows great accuracy with some misclassification. With some uncertainty in light and moderate classes, ~\fref{fig:rockcf} (Rockburst) shows correct estimates for normal and heavy classes, hence highlighting pragmatic constraints. 

\begin{figure}[!t]
\centering
\includegraphics[width=0.4\textwidth,height=5cm]{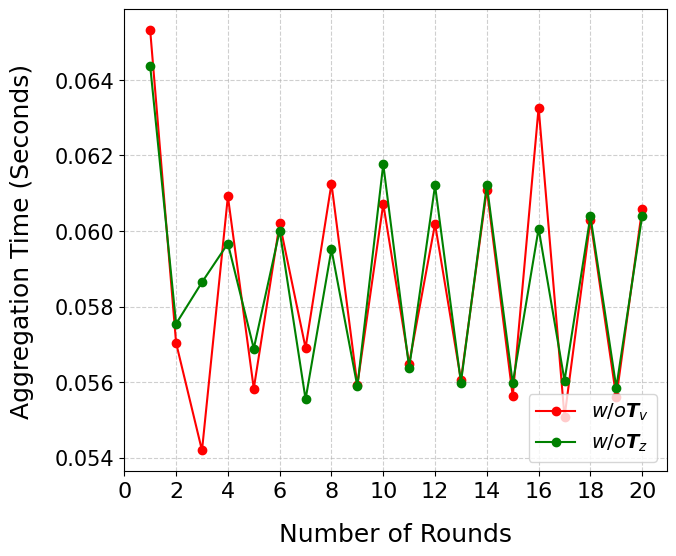}
\caption{Effect of $\boldsymbol{T}_v$ and $\boldsymbol{T}_z$ on aggregation time for Rockburst classification dataset. } 
\label{fig:Tv_Tz} 
\end{figure}
\begin{table}[!t]
\caption{\fontsize{8.9 pt}{11pt}\selectfont Comparison of MUD-HoG vs. MineDetect using FPR detection performance of {\em Sign Flipping} attacks over 10 rounds for the Rockburst classification dataset with attacker Ids 0, 10, 11, and 12.}
\label{tab:sign_flip_detection_rock}
\centering
\resizebox{\columnwidth}{!}{%
\begin{tabular}{c|c|c|c|c}
 \toprule
\multirow{2}{*}{\textbf{Round}} 
& \multicolumn{2}{c|}{\textbf{Detected Sign-Flipping Attacker}}                        
& \multicolumn{2}{c}{\textbf{FPR}} \\ \cline{2-5} 
& MUD-HoG & \multicolumn{1}{c|}{\bf MineDetect (ours)}  
& MUD-HoG & \multicolumn{1}{c}{\bf MineDetect (ours)} \\ \hline
1  & NULL                         & 0,11,12                   & 0\%           & 0\%          \\ \hline
2  & NULL                         & 0, 2, 37, 10, 11, 12, 18, 20 & 0\%           & 11.11\%      \\ \hline
3  & NULL                         & 0,11,3,12                 & 11.11\%       & 2.78\%       \\ \hline
4  & 0, 2, 37, 10, 11, 12, 18, 20 & 0,3,10,11,12               & 11.11\%       & 2.78\%       \\ \hline
5  & 0, 2, 37, 10, 11, 12, 18, 20 & 0,3,10,11,12               & 11.11\%       & 2.78\%       \\ \hline
6  & 0, 2, 37, 10, 11, 12, 18, 20 & 0,3,10,11,12               & 11.11\%       & 2.78\%       \\ \hline
7  & 0, 2, 37, 10, 11, 12, 18, 20 & 0,3,10,11,12               & 11.11\%       & 2.78\%       \\ \hline
8  & 0, 2, 37, 10, 11, 12, 18, 20 & 0,11,3,12                  & 11.11\%       & 2.78\%       \\ \hline
9  & 0, 2, 37, 10, 11, 12, 18, 20 & 0,11,3,12                  & 11.11\%       & 2.78\%       \\ \hline
10 & 0, 2, 37, 10, 11, 12, 18, 20 & 11,3,4,12                  & 11.11\%       & 5.56\%       \\  
\bottomrule
\end{tabular}%
}\vspace{-1mm}
\end{table}

\begin{table}[!t]
\caption{Comparison of MUD-HoG vs. MineDetect using FPR detection performance of {\em Sign Flipping} attacks over 10 rounds for the MNIST dataset with attacker Ids 0, 10, and 11.}
\label{tab:sign_flip_detection_MNIST}
\centering
\resizebox{\columnwidth}{!}{
\begin{tabular}{c|c|c|c|c}
 \toprule
\multirow{2}{*}{\textbf{Round}} 
& \multicolumn{2}{c|}{\textbf{Detected Sign Flip Attacker}} 
& \multicolumn{2}{c}{\textbf{FPR}} \\ \cline{2-5} 
& MUD-HoG & \multicolumn{1}{c|}{\bf MineDetect (ours)} 
& MUD-HoG & \multicolumn{1}{c}{\bf MineDetect (ours)} \\ \hline
1  & NULL                        & 0, 2, 10, 11, 18, 20       & NULL         & 8.11\%       \\ \hline
2  & NULL                        & 0,10,11                    & NULL         & 0\%          \\ \hline
3  & NULL                        & 0,10,11                    & NULL         & 0\%          \\ \hline
4  & 0, 2, 10, 11, 18, 20         & 0,10,11                    & 8.11\%       & 0\%          \\ \hline
5  & 0, 2, 10, 11, 18, 20         & 0,18,10,11                 & 8.11\%       & 2.70\%       \\ \hline
6  & 0, 2, 10, 11, 18, 20         & 0,10,11                    & 8.11\%       & 0\%          \\ \hline
7  & 0, 2, 10, 11, 18, 20         & 0,10,11                    & 8.11\%       & 0\%          \\ \hline
8  & 0, 2, 10, 11, 18, 20         & 0,10,11                    & 8.11\%       & 0\%          \\ \hline
9  & 0, 2, 10, 11, 18, 20         & 0,10,11                    & 8.11\%       & 0\%          \\ \hline
10 & 0, 2, 10, 11, 18, 20         & 0,10,11                    & 8.11\%       & 0\%          \\  
\bottomrule
\end{tabular}\vspace{-3mm}
}
\end{table}

\noindent{\bf Effect of $\boldsymbol{T}_v$ and $\boldsymbol{T}_z$ on Aggregation Time.}
Aggregation time refers to the time it takes to combine the locally computed updates from all mines into a global model. MineDetect has used two dynamic thresholds $\boldsymbol{T}_v$ and $\boldsymbol{T}_z$ to detect \textit{additive noise} attackers in Algorithm~\ref{algorithm3}. In~\fref{fig:Tv_Tz}, the red curve shows the MineDetect aggregation time with only (w/o) $\boldsymbol{T}_v$, and the green curve shows it  w/o $\boldsymbol{T}_z$. The oscillatory pattern in the aggregation time implies that both thresholds dynamically contribute to the filtering and exclusion of adversarial updates, resulting in the variation of computational effort across global rounds.  Nevertheless, the overall aggregation time remains consistent, with minor peaks observed for $\boldsymbol{T}_v$.  This is due to the fact that $\boldsymbol{T}_v$ is contingent upon the variance of local averages from the previous five rounds, while $\boldsymbol{T}_z$ only depends on the magnitude of local averages.

\begin{table}[!t]
\caption{Comparison of MUD-HoG vs. MineDetect using FPR detection performance of {\em Additive Noise} attacks over 10 rounds for the MNIST dataset with attacker Ids  24,25,26, and 27.}
\label{tab:additive_noise_detection_MNIST}
\centering
\resizebox{\linewidth}{!}{
\begin{tabular}{c|cc|cc}
 \toprule
 & \multicolumn{2}{c|}{\textbf{Detected Additive Noise Attacker}} & \multicolumn{2}{c}{\textbf{FPR}} \\ \cline{2-5} 
\multirow{-2}{*}{\textbf{Round}} & \multicolumn{1}{c|}{\bf MUD-HoG} & \bf MineDetect (ours) & \multicolumn{1}{c|}{\bf MUD-HoG} & \bf MineDetect (ours) \\ \hline
1  & \multicolumn{1}{c|}{NULL} & 24,25,26,27 & \multicolumn{1}{c|}{NULL} & 0\% \\ \hline
2  & \multicolumn{1}{c|}{NULL} & 24,25,26,27 & \multicolumn{1}{c|}{NULL} & 0\% \\ \hline
3  & \multicolumn{1}{c|}{NULL} & 24,25,26,27 & \multicolumn{1}{c|}{NULL} & 0\% \\ \hline
4  & \multicolumn{1}{c|}{NULL} & 24,25,26,27 & \multicolumn{1}{c|}{NULL} & 0\% \\ \hline
5  & \multicolumn{1}{c|}{24,25,26,27} & 24,25,26,27 & \multicolumn{1}{c|}{0\%} & 0\% \\ \hline
6  & \multicolumn{1}{c|}{24,25,26,27} & 24,25,26,27 & \multicolumn{1}{c|}{0\%} & 0\% \\ \hline
7  & \multicolumn{1}{c|}{24,25,26,27} & 24,25,26,27 & \multicolumn{1}{c|}{0\%} & 0\% \\ \hline
8  & \multicolumn{1}{c|}{24,25,26,27} & 24,25,26,27 & \multicolumn{1}{c|}{0\%} & 0\% \\ \hline
9  & \multicolumn{1}{c|}{24,25,26,27} & 24,25,26,27 & \multicolumn{1}{c|}{0\%} & 0\% \\ \hline
10 & \multicolumn{1}{c|}{24,25,26,27} & 24,25,26,27 & \multicolumn{1}{c|}{0\%} & 0\% \\ \bottomrule
\end{tabular}
}
\end{table}

\begin{table}[!t]
\caption{Comparison of MUD-HoG vs. MineDetect using FPR detection performance of {\em Unreliable Mines} over 10 rounds for the MNIST dataset with unreliable Ids 2, 18, 20, and 37.}
\label{tab:unreliable_mine_detection_MNIST}
\centering
\resizebox{\linewidth}{!}{
\begin{tabular}{c|cc|cc}
 \toprule
 & \multicolumn{2}{c|}{\textbf{Detected Unreliable IDs}} & \multicolumn{2}{c}{\textbf{FPR}} \\ \cline{2-5} 
\multirow{-2}{*}{\textbf{Round}} & \multicolumn{1}{c|}{\bf MUD-HoG} & \bf MineDetect (ours) & \multicolumn{1}{c|}{\bf MUD-HoG} & \bf MineDetect (ours) \\ \hline
1  & \multicolumn{1}{c|}{NULL} & 0, 10, 11, 12, 24, 25, 26, 27 & \multicolumn{1}{c|}{NULL} & 22.22\% \\ \hline
2  & \multicolumn{1}{c|}{NULL} & 2, 20, 37 & \multicolumn{1}{c|}{NULL} & 0\% \\ \hline
3  & \multicolumn{1}{c|}{NULL} & 2, 18, 20, 37 & \multicolumn{1}{c|}{NULL} & 0\% \\ \hline
4  & \multicolumn{1}{c|}{6, 8, 17, 18, 20, 23, 24, 25, 26} & 2, 18, 20, 37 & \multicolumn{1}{c|}{19.44\%} & 0\% \\ \hline
5  & \multicolumn{1}{c|}{6, 8, 17, 18, 20, 23, 24, 25, 26} & 2, 18, 37 & \multicolumn{1}{c|}{19.44\%} & 0\% \\ \hline
6  & \multicolumn{1}{c|}{6, 8, 17, 18, 20, 23, 24, 25, 26} & 2, 18, 20, 37 & \multicolumn{1}{c|}{19.44\%} & 0\% \\ \hline
7  & \multicolumn{1}{c|}{2, 18, 20, 24, 25, 26, 29} & 2, 20, 37 & \multicolumn{1}{c|}{11.11\%} & 0\% \\ \hline
8  & \multicolumn{1}{c|}{2, 18, 20, 24, 25, 26, 29} & 2, 20, 37 & \multicolumn{1}{c|}{11.11\%} & 0\% \\ \hline
9  & \multicolumn{1}{c|}{2, 18, 20, 24, 25, 26, 29} & 2, 18, 20, 37 & \multicolumn{1}{c|}{11.11\%} & 0\% \\ \hline
10 & \multicolumn{1}{c|}{2, 18, 20, 24, 25, 26, 29} & 2, 20, 37 & \multicolumn{1}{c|}{11.11\%} & 0\% \\  
\bottomrule
\end{tabular}
}
\end{table}

\vspace{1mm}
\noindent{\em  {\bf 6.2.3  Ablation Study}} 

\noindent To assess the efficacy of MineDetect, we have performed an ablation study, isolating its distinct detection algorithm to highlight the independent performance of each detection algorithm.
Table~\ref{tab:sign_flip_detection_rock} and~\ref{tab:sign_flip_detection_MNIST} present the detection outcomes of Algorithm \ref{algorithm2} and comparing with MUD-HoG. In the Rockburst classification dataset (Table~\ref{tab:sign_flip_detection_rock}), featuring four attackers, MineDetect has a markedly lower FPR of 2.78\% in contrast to MUD-HoG's 11.11\%. In the MNIST dataset (Table~\ref{tab:sign_flip_detection_MNIST}), which has three attackers, MineDetect attains a lower FPR of 0\% from Round 2 onwards, in contrast to MUD-HoG's persistent FPR of 8.11\% from Round 4. This decrease emphasizes that MineDetect runs effectively across different amounts of attackers. Similarly, only Algorithm \ref{algorithm3} is executed, and Table~\ref{tab:additive_noise_detection_MNIST} presents the comparison of FPR with MUD-HoG. MineDetect effectively detects all attackers from the first round with a 0\% FPR, whereas MUD-HoG identifies attackers starting from Round 5.  Likewise, Table~\ref{tab:unreliable_mine_detection_MNIST} compares the FPR for detecting \textit{unreliable mines}. MineDetect successfully detects them from the second round with a 0\% FPR, while MUD-HoG exhibits a variable FPR between 19.44\% and 11.11\%. These results confirm that MineDetect's history-aware and variance-based algorithms enable independently reliable detection with low FPR across multiple datasets and attacker scenarios.

\section{Conclusion and Future Work}
This paper proposes MineDetect, a history-aware FL defense framework designed to detect threats in underground mining operations. MineDetect effectively isolates  sign-flipping and additive noise attacks, while also identifying and mitigating unreliable mines that contribute with poor-quality data. For each global round, MineDetect keeps track of the local averages of individual mines that capture a historical behavior of updated parameters. In addition, the global average is calculated as a solid reference point by taking the mean of all local averages from each mine in a specific round. These two factors strengthen MineDetect's capability to distinguish normal mines from compromised mines. Experimental results, including application-specific data of the Rockburst classification dataset, demonstrate MineDetect's robustness, maintaining high accuracy and low false positive rates (FPR) even as attacker percentages increase. MineDetect improves accuracy by 5.94\% and reduces the FPR by 8.2\%, highlighting its strong performance. 
Ablation studies further confirm the importance of each detection algorithm to enhance the overall strength of the framework.

In the future, we plan to improve MineDetect by incorporating secure aggregation techniques to safeguard sensitive mining data. Additionally, we aim to consider more sophisticated attack types. We also plan to make our own dataset collecting from experimental mine located in our campus.

\bibliographystyle{IEEEtran}
\renewcommand{\baselinestretch}{.9}
\small{
\bibliography{_references.bib}
}
\end{document}